\documentclass{jpconf}

\usepackage{epsfig}
\usepackage{graphicx}
\usepackage{hhline}
\usepackage{color}

\definecolor{darkblue}{rgb}{0,0,0.5}
\definecolor{green}{rgb}{0,.6,0}
\definecolor{SkyBlue}        {named}{SkyBlue}

\begin{document}

\title{Results from the $N^*$ Program at JLab}

\author{I. Aznauryan$^{1,2}$, V.D. Burkert$^1$, 
T.-S. H. Lee$^{3,4}$, V. Mokeev$^{1,5}$}

\address{$^1$Thomas Jefferson National
Accelerator Facility, Newport News, Virginia 23606, USA}
\address{$^2$ Yerevan Physics Institute, Yerevan 375036, Armenia}
\address{$^3$Physics Division, Argonne National Laboratory, Argonne,
 Illinois 60439, USA }
\address{$^4$ Excited Baryon Analysis Center,
Thomas Jefferson National Accelerator Facility, Newport News, 
Virginia 23606, USA}
\address{$^5$ Skobeltsyn Nuclear Physics Institute at 
Moscow State University, 119899,
Moscow, Russia}
\ead{aznaury@jlab.org, burkert@jlab.org, 
lee@phy.anl.gov, mokeev@jlab.org}

\begin{abstract}
We discuss the results on the fundamental
degrees of freedom underlying the nucleon excitation spectrum and
how they evolve as the resonance transitions are investigated with
increasingly better  space-time resolution of the electromagnetic probe.
Improved photocouplings for a number
of resonant states, those for the N(1720)P$_{13}$ being significantly
changed, have been determined and  
entered into the
2008 edition of the RPP. 
Strong sensitivity
to the N(1900)P$_{13}$ state, that is listed now as a 2-star
state in the same edition of RPP, has been observed
in $K\Lambda,~K\Sigma $ photoproduction.
None of the earlier
observations of a
$\Theta^+_5(1540)$
was confirmed in a series of three JLab high statistics
dedicated measurements, and stringent upper limits
on production cross sections
were placed in several channels.
For the four lowest excited
states, the
$\Delta$(1232)P$_{33}$, N(1440)P$_{11}$,
N(1520)D$_{13}$, and
N(1535)S$_{11}$,
the transition amplitudes have been measured in a wide
range in photon virtuality $Q^2$.
The amplitudes for the $\Delta$ show the importance of
the pion-cloud contribution 
and don't show any sign of approaching the pQCD regime for
$Q^2<7~$GeV$^2$.  For
the Roper resonance, N(1440)P$_{11}$, the data provide strong evidence
for this
state as a predominantly radial excitation of the nucleon as a 3-quark
ground state. 
For the N(1535)S$_{11}$,
comparison of the results 
extracted from $\pi$ and
$\eta$ photo- and electroproduction data
allowed one to specify the branching ratios
of this state to the $\pi N$ and $\eta
N$ channels; they will
enter into the
2010 edition of the RPP.
Measured for the first time, the longitudinal transition amplitude
for the N(1535)S$_{11}$
became a challenge for quark models
and can be indicative of
large meson-cloud contributions or
alternative representations of this state.
The N(1520)D$_{13}$ clearly shows the rapid
changeover from
helicity-3/2 dominance at the real photon point to
helicity-1/2
dominance at $Q^2 > 0.5~$GeV$^2$, confirming a long-standing prediction
of the constituent quark model.  The search for undiscovered but
predicted
states continues to be pursued with a vigorous experimental program.
While recent data from JLab and elsewhere provide intriguing hints of
new states, final conclusions will have to wait for the results of the
broad experimental effort currently underway with CLAS, and subsequent
analyses involving the EBAC at JLab.
\end{abstract}

\section{Introduction}
Over 98\% of the mass of the visible universe is made of nucleons. They are at
the core of atoms and nuclei in stars that provide the energy that heat the
planets and allows for life to exist on at least one of them, our earth.
Understanding their internal structure has been at the center of nuclear
and particle physics for decades. Similar to atomic
nuclei, nucleons are complex systems of confined quarks
and gluons and exhibit
characteristic spectra of excited states that encode information on 
symmetry properties of baryonic matter. 
Highly excited nucleon states ($N^*$) are sensitive to details of
quark confinement, which is poorly understood within
Quantum Chromodynamics (QCD), the fundamental theory of strong interactions.
Measurements  of excited nucleon states are needed
to come to a better understanding of how confinement works in nucleons. 
The clustering of quarks in the nucleon can lead to an excitation spectrum 
with fewer states than if quarks do not cluster. 
The $N^*$ states couple strongly to the meson-baryon continuum to form
nucleon resonances of characteristic masses and decay widths, and can be most
effectively investigated by using meson production reactions on the nucleon.
With the high precision electron beam and the CEBAF Large 
Acceptance Spectrometer 
(CLAS) in Hall B, Jefferson Laboratory (JLab) has developed an excited 
baryon program and has 
made
important contributions in this direction. 

The excited baryon program at JLab has two main objectives.
The first one is to provide information from electromagnetic probes
to improve our understanding of the systematics of the $N^*$ spectrum
and the nature of the effective degrees of freedom in low
energy QCD.  The second objective
is to measure the resonance transition form factors from the nucleon ground
state to the excited baryon states.
These measurements probe the internal structure of excited states
and provide information about the confining forces of the 3-quark system
and the spatial distribution of quarks in the transverse plane.

\begin{figure}[t]
 \includegraphics[height=.34\textheight]{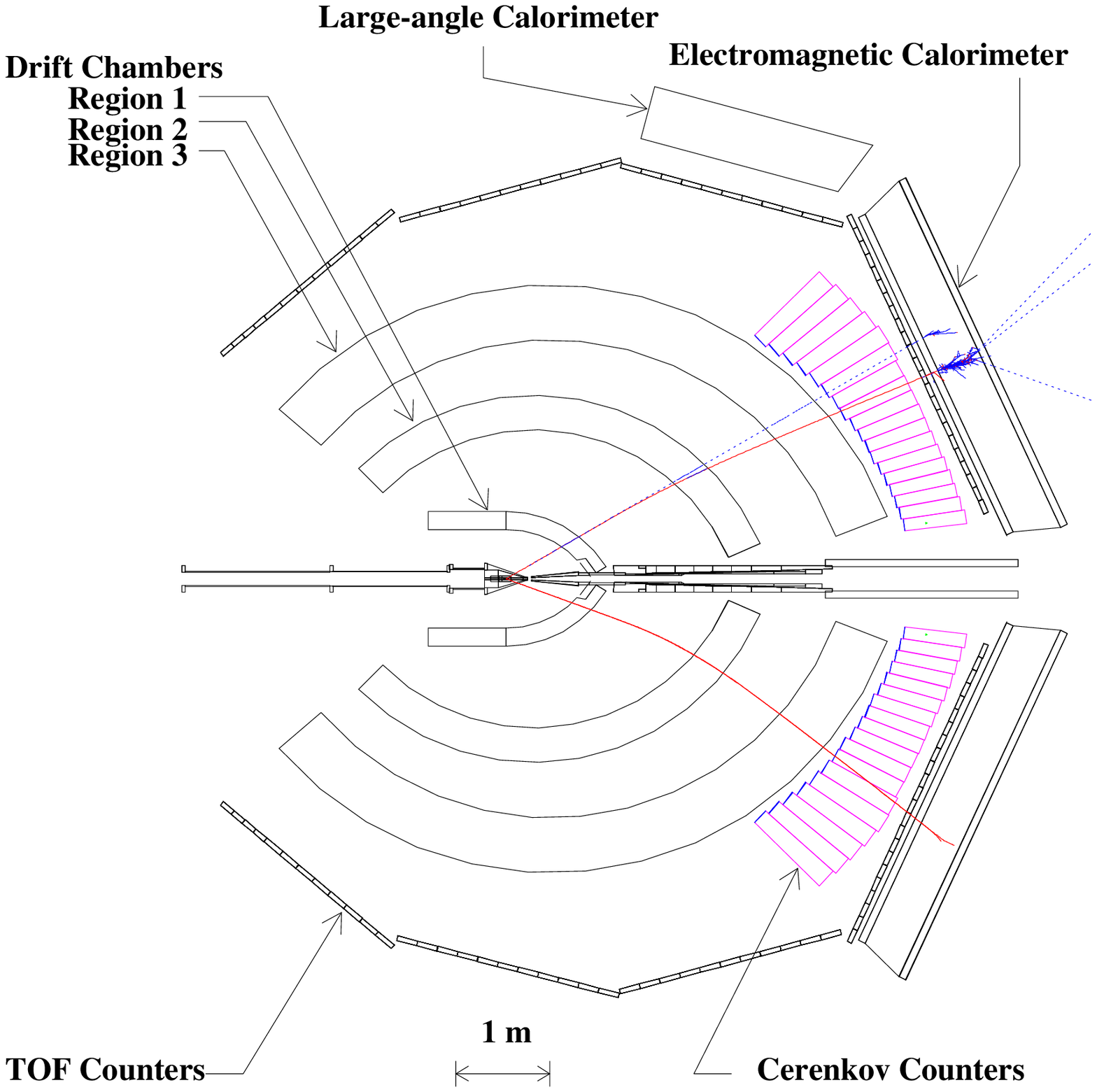}
  \includegraphics[height=.34\textheight]{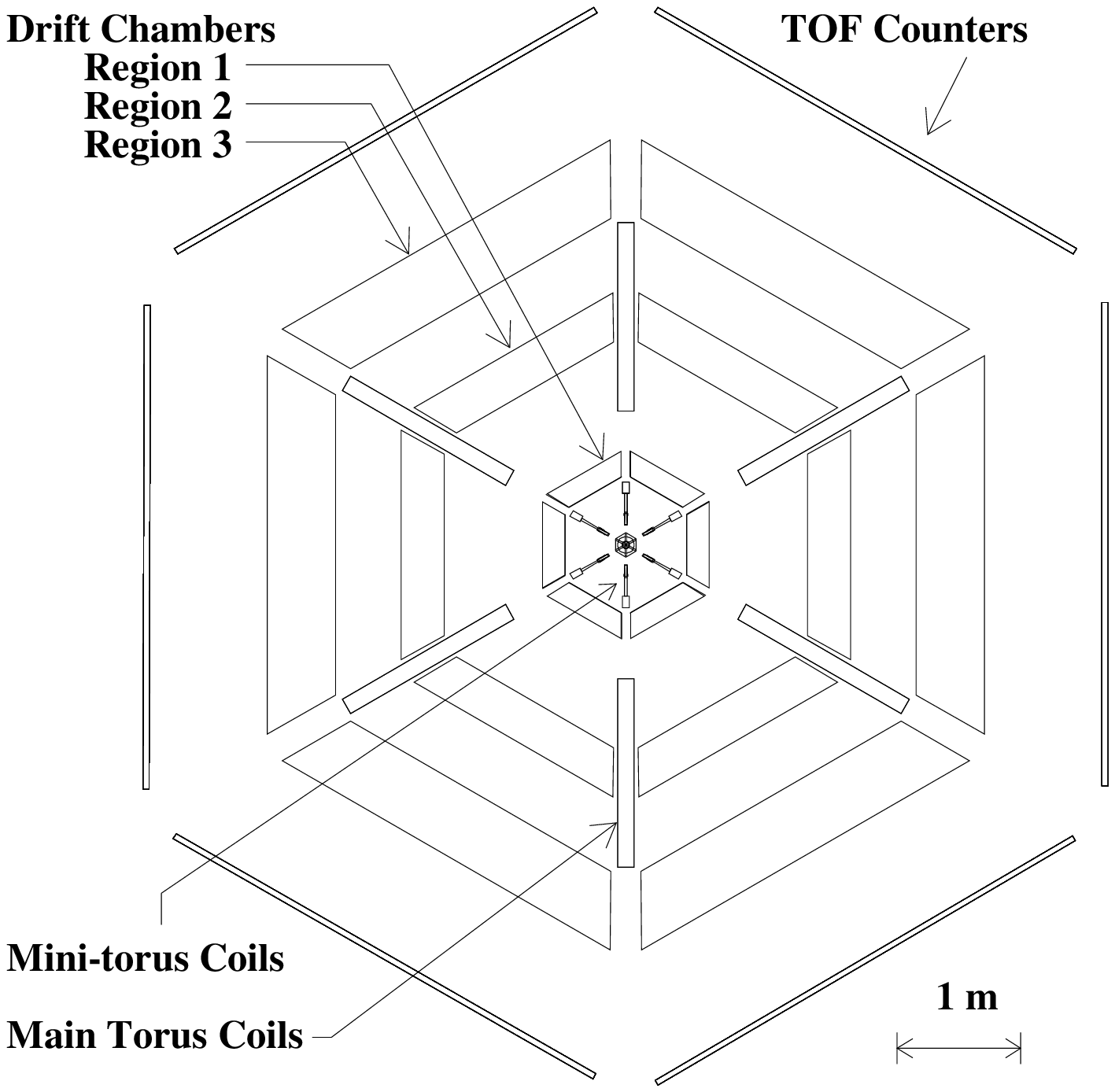}
  \caption{The CEBAF Large Acceptance Spectrometer (CLAS). A cut along
the beamline (with the beam incident from the left) is shown on the l.h.s.,
while the r.h.s. shows a cut perpendicular to the beam. A superconducting 
toroidal magnet provides
the magnetic analysis in six independent sectors. Each of the six sectors 
is independently 
instrumented with tracking chambers and with detectors for particle 
identification.}
\label{fig:clas}
\end{figure}

The experimental program makes wide use of the CLAS detector~\cite{mecking}, 
which is depicted in Fig. \ref{fig:clas}. 
CLAS provides particle detection and identification, as well as 
momentum analysis in a polar angle range from 8$^\circ$ to 140$^\circ$. 
For operation of energy-tagged photon beams, the 
photon energy tagging facility provides photons with an energy resolution of 
$\frac{\sigma(E)}{E} = 10^{-3}$, and covers an energy range from 20\% to 95\% of the 
incident electron beam energy. Other equipment essential for the $N^*$ 
program includes the coherent 
bremsstrahlung facility that has been used to produce linearly polarized photons with 
polarizations up to 90\%. There are also two frozen-spin polarized 
targets, one using butanol as target material for polarized protons (FROST), 
and the other using HD as an efficient neutron target (HD-Ice). FROST has been operated 
successfully in longitudinal polarization mode, in conjunction with linearly and circularly
polarized photons,  and is planned to be used in 
transverse polarization mode in 2010. HD-Ice is planned to be used 
as a polarized neutron target 
in 2010/2011. In addition, the highly polarized electron beam, $P_e = 0.85$, 
generates circularly polarized photons when scattered off an amorphous 
radiator. The polarization transfer is maximum at the highest photon energies.

CLAS is also being employed for the measurement of electroproduction of 
single and double pions,
and other pseudoscalar mesons, e.g. $K^+$, $K^-$, $K_s^0$. Some of 
these experiments 
make use of a dynamically polarized 
nucleon target~\cite{Keith}. Some experiments, focusing on specific 
kinematics, make use of the 
magnetic spectrometer 
setups in Hall A, using  
proton recoil polarimeters for the measurement of single and 
double polarization observables, 
and in Hall C to study high $Q^2$ pion and eta production.
The list of experiments approved to study
$N^*$ excitations 
or search for new states is given in Table \ref{tab:experiments}.

In addition to analyzing the  electromagnetic meson production data,
it is important to interpret the extracted $N^*$ parameters in terms of QCD.
The Excited Baryon Analysis Center (EBAC) was established at JLab in 2006
to address this problem. This effort is making rapid progress by making
a dynamical coupled-channels analysis of the world data of
both the hadronic and electromagnetic meson production data.
 In addition, effort has been made to make contact with hadron structure 
calculations, in particular the LQCD effort at JLab.

In Sections 2 and 3, the results from analyzing the JLab photoproduction and
electroproduction data will be described.
The progress and status of EBAC will be
given in Section 4. Section 5 is devoted to the discussion of
the future prospects of the $N^*$ program at 6  GeV and 
of  the possibilities
with the 12-GeV upgrade.

\begin{table}[t]
\begin{tabular}{|lc|lc|}
\hline
&&&\\
$p(\gamma,K)X$&\cite{E89004}&
$p(e,e'K^+K^-)p$,$p(e,e'K^+\pi^-)X$&\cite{E89043}\\
$p(\gamma,\eta,\eta')p$&\cite{E91008}&
$p(e,e'\pi)N$&\cite{E91002}\\
$D(\gamma,\eta),~D(\gamma,\eta')$&\cite{E94008}&
$\vec{p}(\vec{e},e'p)\pi^0$&\cite{E91011}\\
$p(\gamma,\pi)N$&\cite{E94103}&
$p(e,e'\omega)p$&\cite{E91024}\\
$p(\vec{\gamma},\omega)p$&\cite{E99013}&
$p(e,e'\pi^+\pi^-)p$&\cite{E93006}\\
$\vec{\gamma}\vec{p}\rightarrow K^+\Lambda,K^+\Sigma,K^0\Sigma^+$&\cite{E02112}&
$\vec{p}(\vec{e},e'\pi)n$&\cite{E93036}\\
$\vec{p}({\vec\gamma},\pi^+)n,\vec{p}({\vec\gamma},p)\pi^0$&\cite{E03105}&
$p(\vec{e},e'p)\pi^0$, $p(\vec{e},e'\pi^+)n$&\cite{E94003}\\
$\vec{p}({\vec\gamma},\eta)p$&\cite{E05012}&
$p(e,e'p)\pi^0$&\cite{E94014}\\
$\vec{p}(\vec{\gamma},\pi^+\pi^- p)$&\cite{E06013}&
$p(e,e'K^{+})\Lambda,\Sigma$&\cite{E99006}\\
$n(\vec{\gamma},K\Lambda)$&\cite{E06103}&
$p(e,e'\pi^0)p$,$p(e,e'\pi^+)n$&\cite{E99107}\\
$p(e,e'\pi)N$&\cite{E89037}&
$p(e,e'\pi^+\pi^-p)$&\cite{E99108}\\
$p(e,e'\pi^+)n$, $p(e,e'p)\pi^0$, $n(e,e'\pi^-)p$&\cite{E89038}&
$p(\vec{e},e'K,\vec{\Lambda},\vec{\Sigma})$&\cite{E00112}\\
$p(e,e'p)\eta$&\cite{E89039}&
$p(e,e'p)\pi^0,\eta$&\cite{E01002}\\
$p(\vec{e},e'p)\pi^0$, $p(\vec{e},e'\pi^+)n$&\cite{E89042}&&\\
&&&\\
\hline
\end{tabular}
\caption{\label{tab:experiments}
Experiments at JLab that are part of the $N^*$ program. 
}
\end{table}

\section{Results from Photoproduction Experiments}
\subsection{Photoproduction of pseudoscalar mesons}
\begin{figure}
\psfig{file=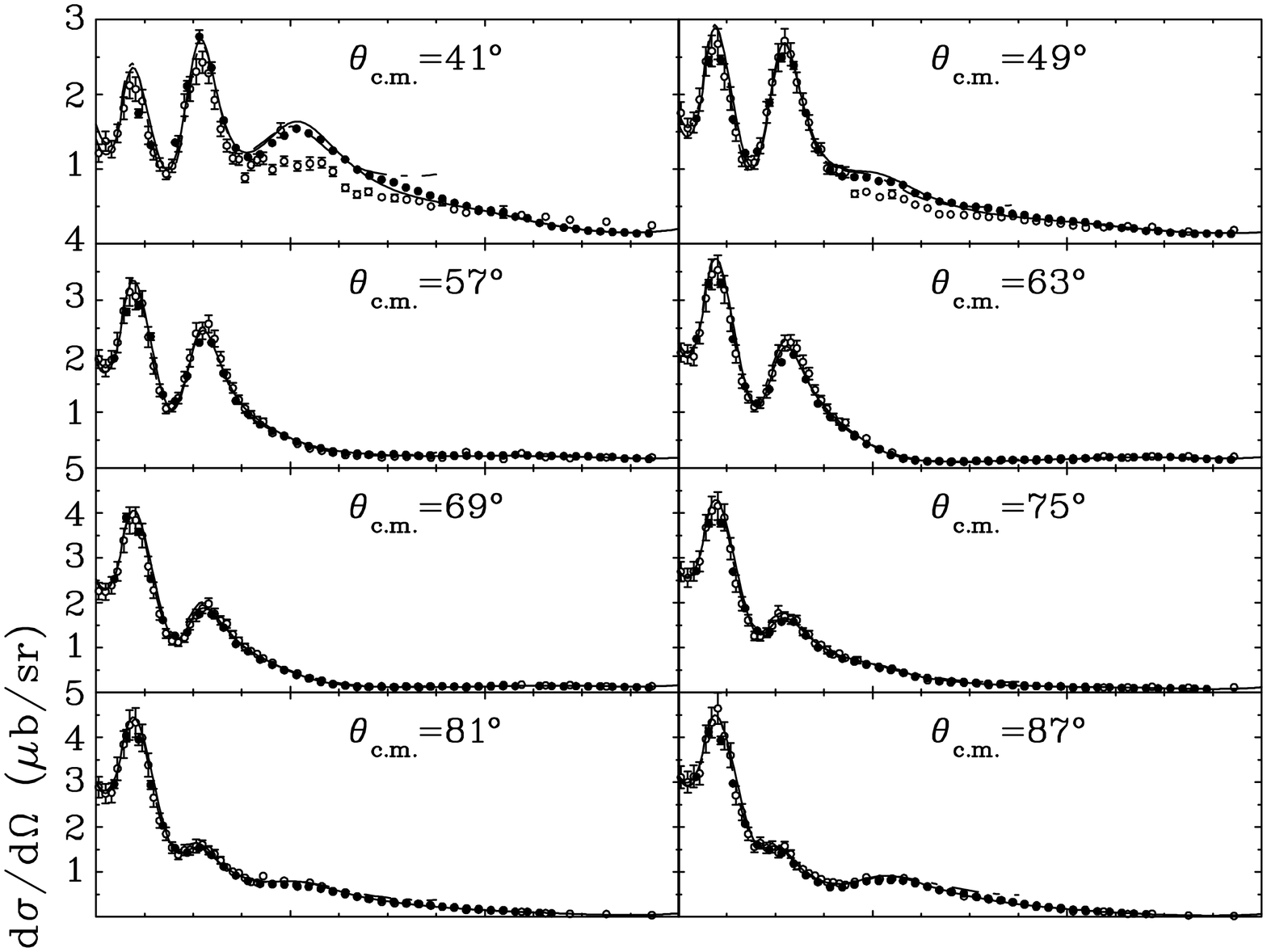, height=2.5in, angle=90}
\psfig{file=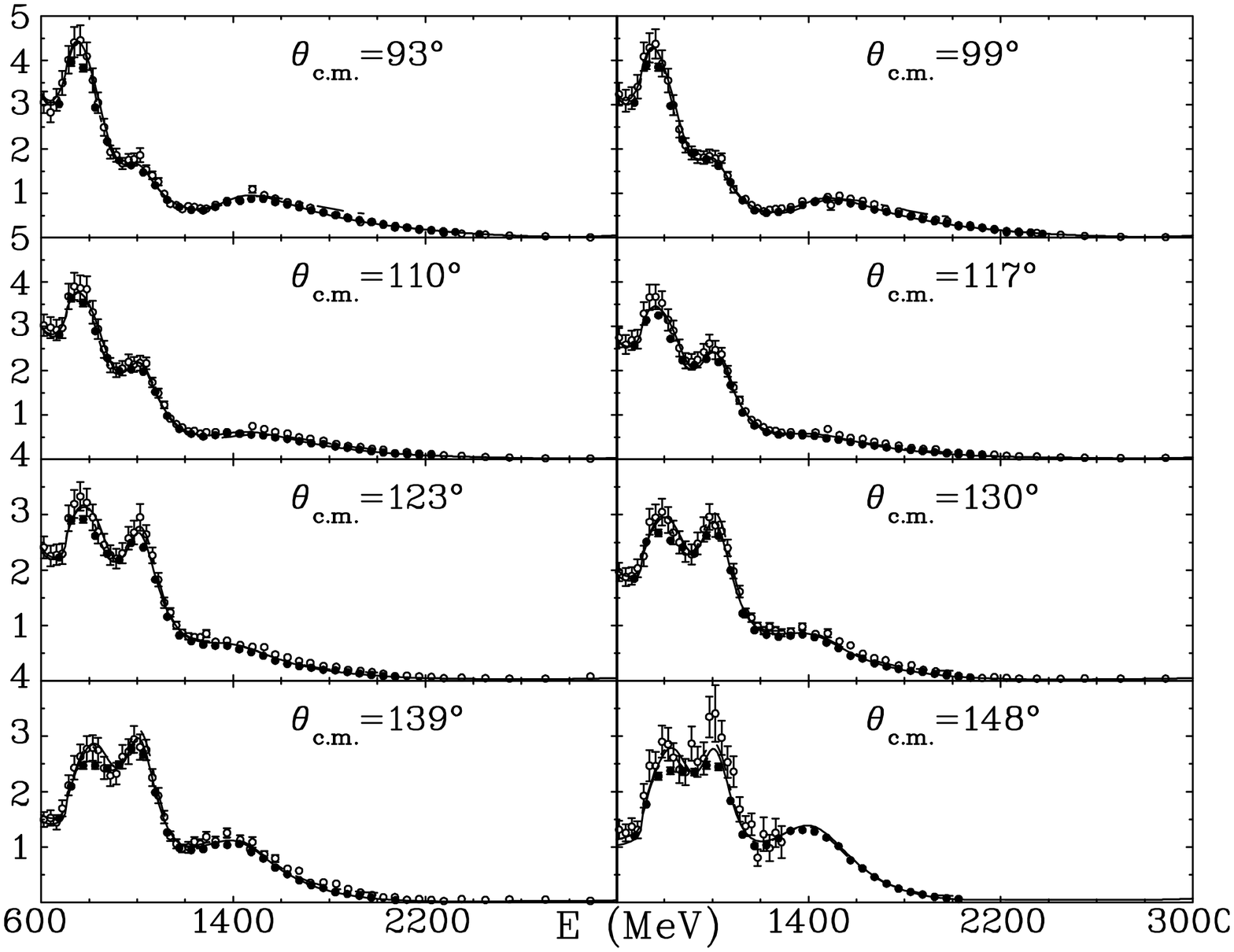, height=2.5in, angle=90}
\caption{Differential cross sections from CLAS 
for $\gamma p \rightarrow n\pi^+$ 
\cite{dugger09} 
vs. photon energy $E$ ($W^2=m_p^2+2m_p E$). 
Solid and open circles show, respectively, the CLAS 
and previous data.
New measurements \cite{dugger09}
have improved the world data base significantly in energy 
range and polar angle coverage.
Curves show the results of the SAID FA08 solution. }
\label{fig:pipl-cs}
\end{figure}

Precise differential cross sections on proton targets for final states 
$p\pi^0$~\cite{dugger07}, $n\pi^+$~\cite{dugger09}, $p\eta$~\cite{dugger02}, 
$p\eta^\prime$~\cite{dugger06}, $K^+\Lambda$~\cite{mcnabb04,brad06}, and 
$K^+\Sigma^0$~\cite{mcnabb04,brad06} have been measured. 
As an example, Fig.~\ref{fig:pipl-cs} 
shows differential cross sections for $\gamma p \rightarrow n \pi^+$ for different
hadronic mass $W$. These data, together with the $\pi N \rightarrow \pi N$ 
elastic data and 
the cross sections from the $\gamma p \rightarrow p\pi^0$ channel also measured 
with CLAS,
have been used to determine improved helicity amplitudes for a number of 
resonant states 
with significant couplings to the single pion channel. They have been also 
entered into the 
2008 edition of the Review of Particle Physics (RPP)~\cite{pdg2008}. For 
example, 
the photocoupling of 
the N(1720)P$_{13}$ 
has changed significantly with the new precise data. 
More complex processes such as 
$\gamma p \rightarrow p \pi^+ \pi^-$ are 
also being studied. The latter are particularly sensitive to higher-mass 
nucleon resonances 
that couple dominantly to the $\Delta\pi$ or $N\rho$ isobars with two 
pions in 
the final state.  

In photoproduction of single pions or etas,
partial wave analyses of differential cross sections 
alone result in ambiguous solutions 
for the contributing resonant partial waves. The CLAS $N^*$ 
program is therefore 
aimed at complete or nearly complete measurements for all of these processes. 
Complete information may be obtained using a combination of linearly and circularly 
polarized photon beams, 
measurement of the hyperon recoil polarization, and using longitudinally 
and transversely polarized targets. Single pseudoscalar meson production 
is fully described by 4 complex, 
parity conserving amplitudes, requiring eight \cite{Keaton} well-chosen combinations of 
beam, target, and recoil polarization measurements for an 
unambiguous extraction of the production amplitude. Table~\ref{tab:complete} 
shows the observables 
that can be determined from these measurements. If all possible combinations are 
measured, 16 observables can be extracted. In measurements that involve nucleons 
in the final state with no recoil polarization being measured, 7 independent 
observables 
can be measured directly. In addition, the recoil polarization asymmetry $P$ can 
be inferred 
from the double polarization asymmetry with a linearly polarized beam and 
transverse target 
polarization. 

\begin{figure}
\psfig{file=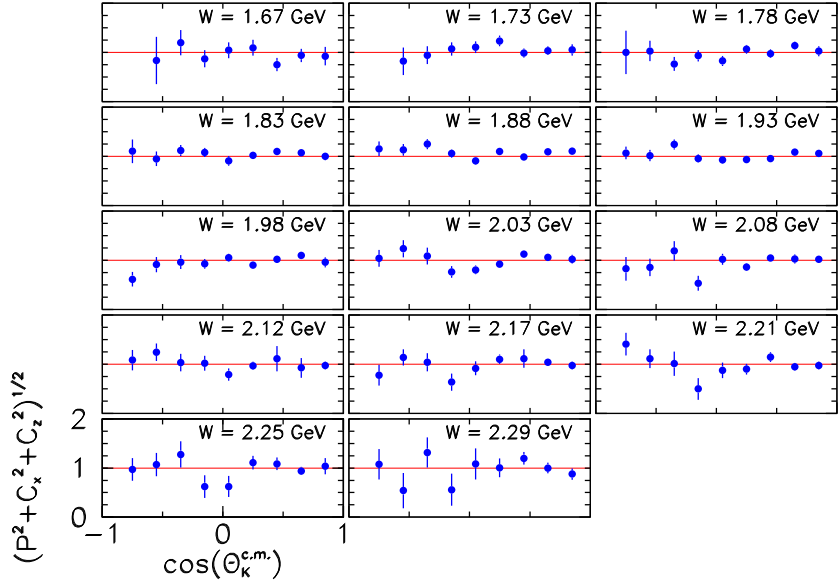, width=5.0in}
\caption{CLAS data \cite{brad07} on 
the polarization observable vector
combining recoil polarization $P$
and transferred polarization components $C_x$ and $C_z$ 
for the $\Lambda$ hyperons produced in 
$\vec{\gamma}+p\rightarrow K^+ +\vec{\Lambda}$.
Lower-left axis scales apply to all plots.
The data show the $\Lambda$ is 100\% polarized.
These data were crucial when fitted simultaneously with the CLAS 
$K^+\Lambda$ and $K^+\Sigma$ differential 
cross section data.  They provided evidence for an 
N(1900)P$_{13}$ candidate state, 
which is required for a good simultaneous fit 
to the cross section and spin transfer data.}
\label{fig:lambda_spin}
\end{figure}

In addition to precise $K\Lambda$ and $K\Sigma$
cross section data, recoil polarization and polarization transfer data have been 
measured~\cite{brad07}. The recoil polarization data in the $K^+\Lambda$ 
sector showed a 
highly unexpected behavior: the spin transfer from the circularly polarized 
photon to the $\Lambda$ 
hyperon is complete, creating the $\Lambda$ hyperon 100\% polarized, as can be seen 
in Fig.~\ref{fig:lambda_spin}. 
The sum of all polarization components $R^2 \equiv P^2 + C_x^2 + C_z^2$, which 
has an upper bound of 1, 
is consistent with $R^2 = 1$ throughout the region covered by the measurement. 
The combined analysis of the CLAS cross section and polarization 
transfer data by the Bonn-Gatchina group~\cite{nikonov08} shows 
strong sensitivity to a 
N(1900)P$_{13}$ candidate state. The decisive ingredient in this 
analysis 
was the CLAS spin transfer 
data. An N(1900)P$_{13}$ state is listed as a 2-star candidate
state in the 2008 edition of the RPP~\cite{pdg2008}. If this 
assignment is further corroborated, 
the existence of an N(1900)P$_{13}$ state will provide strong 
evidence 
against 
the quark-diquark model 
that has no place for such a state in this mass range~\cite{santo2005}. 

\begin{figure}
\begin{center}
 \includegraphics[height=.4\textheight]{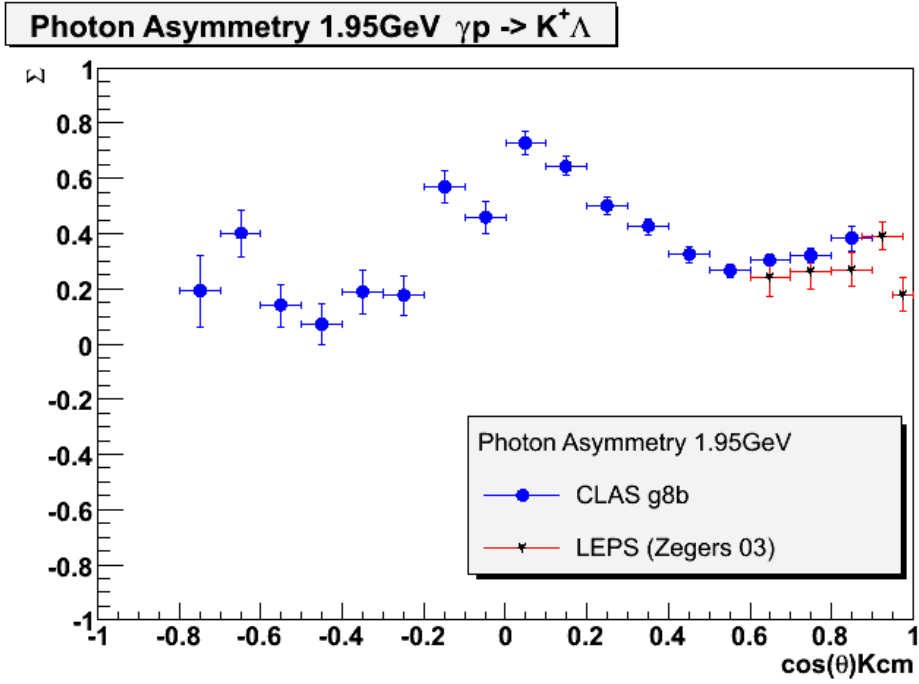}
\end{center}
\caption{Beam asymmetry for  
$\vec{\gamma}p\rightarrow K^+\Lambda$ with 
linearly polarized photons. 
Preliminary CLAS data are shown for   
$W=1.95~$GeV.
The previous data are from GRAAL \cite{Lleres} and LEPS \cite{Zegers}.}
\label{fig:g8b_asymmetry}
\end{figure}

New precise high statistics data with a linearly polarized photon beam on a
proton target are 
becoming also available. Preliminary results for the process 
$\vec{\gamma} p \rightarrow K^+\Lambda$ in just one energy 
bin are shown in Fig.~\ref{fig:g8b_asymmetry}. For the 
partial wave analysis, complete angle coverage in all observables 
is especially important.

\begin{table}[t]
\begin{tabular}{|c|ccc|ccc|ccccccccc|}
\hline
Photon &&T&&&R&&&&T&-&R&&&&\\
\hline
&&&&$x'$&$y'$&$z'$&$x$&
$y$&$z$&$x$&$y$&$z$&$x$&$y$&$z$\\
\hline
&$x$&$y$&$z$&&&&$x$&
$y$&$z$&$x$&$y$&$z$&$x$&$y$&$z$\\
\hline
I&&$T$&&&$P$&&
$T_{x'}$&&$L_{x'}$&&{\color{green}$\Sigma$}&&$T_{z'}$&&$L_{z'}$\\
\hline
II&$H$
&{\color{green}$P$}&$G$&$O_{x'}$&{\color{green}$T$}&$O_{z'}$&
{\color{blue}$L_{z'}$}&{\color{blue}$C_{z'}$}&{\color{blue}$T_{z'}$}
&{\color{blue}$E$}&&{\color{blue}$F$}&{\color{blue}$L_{x'}$}
&{\color{blue}$C_{x'}$}&{\color{blue}$T_{x'}$}\\
\hline
III&$F$&&$E$&$C_{x'}$&&$C_{z'}$&
&{\color{blue}$O_{z'}$}&&{\color{blue}$G$}&&{\color{blue}$H$}&&{\color{blue}$O_{x'}$}&\\
\hline
\end{tabular}
\caption{\label{tab:complete}
Polarization observables in $J^{\pi}=0^-$ meson
photoproduction
that can be extracted from complete experiments using 
all combinations of beam, target, and recoil polarizations. 
I,II,III correspond, respectively, to unpolarized, linearly and circularly
polarized photon beams.
The observables given in black
letters are directly extractable. Other
observables provide additional redundant information:
green letters show the single-polarization observables measured from 
double-polarization asymmetries, blue letters show 
the double-polarization observables measured from 
triple-polarization asymmetries. 
 For example, the target polarization
asymmetry $T$ can be extracted in a single-polarization measurement with unpolarized beam
and a transversely polarized target, and also in a double-polarization measurement with 
a linearly polarized beam and measurement of the recoil polarization $P_{y'}$. This allows 
multiple cross checks of measurements with different systematics.
}
\end{table}

\subsection{Search for new cascade ($\Xi$) baryons}
\begin{figure}
\begin{center}
  \includegraphics[height=.24\textheight]{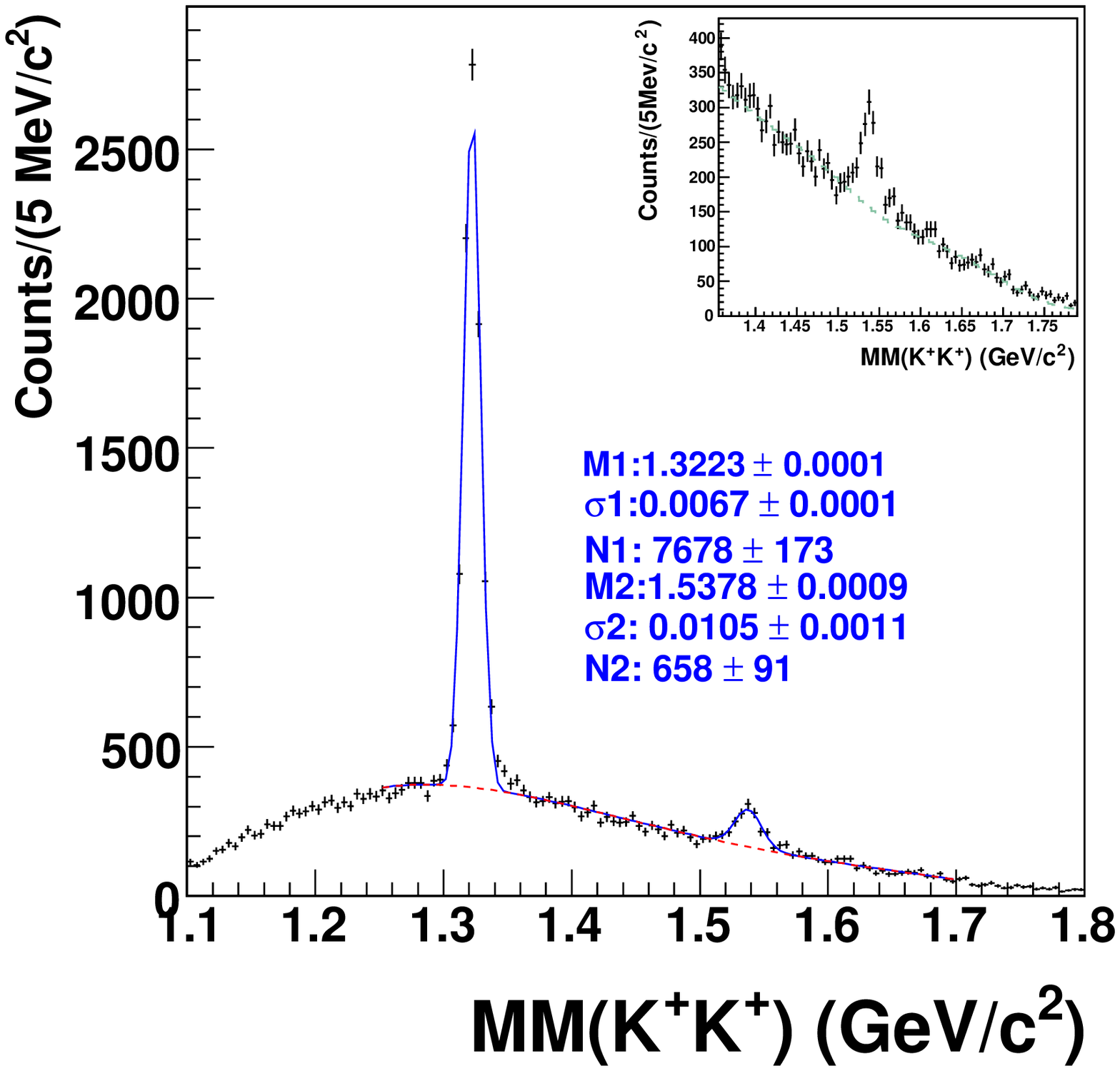}
  \includegraphics[height=.24\textheight]{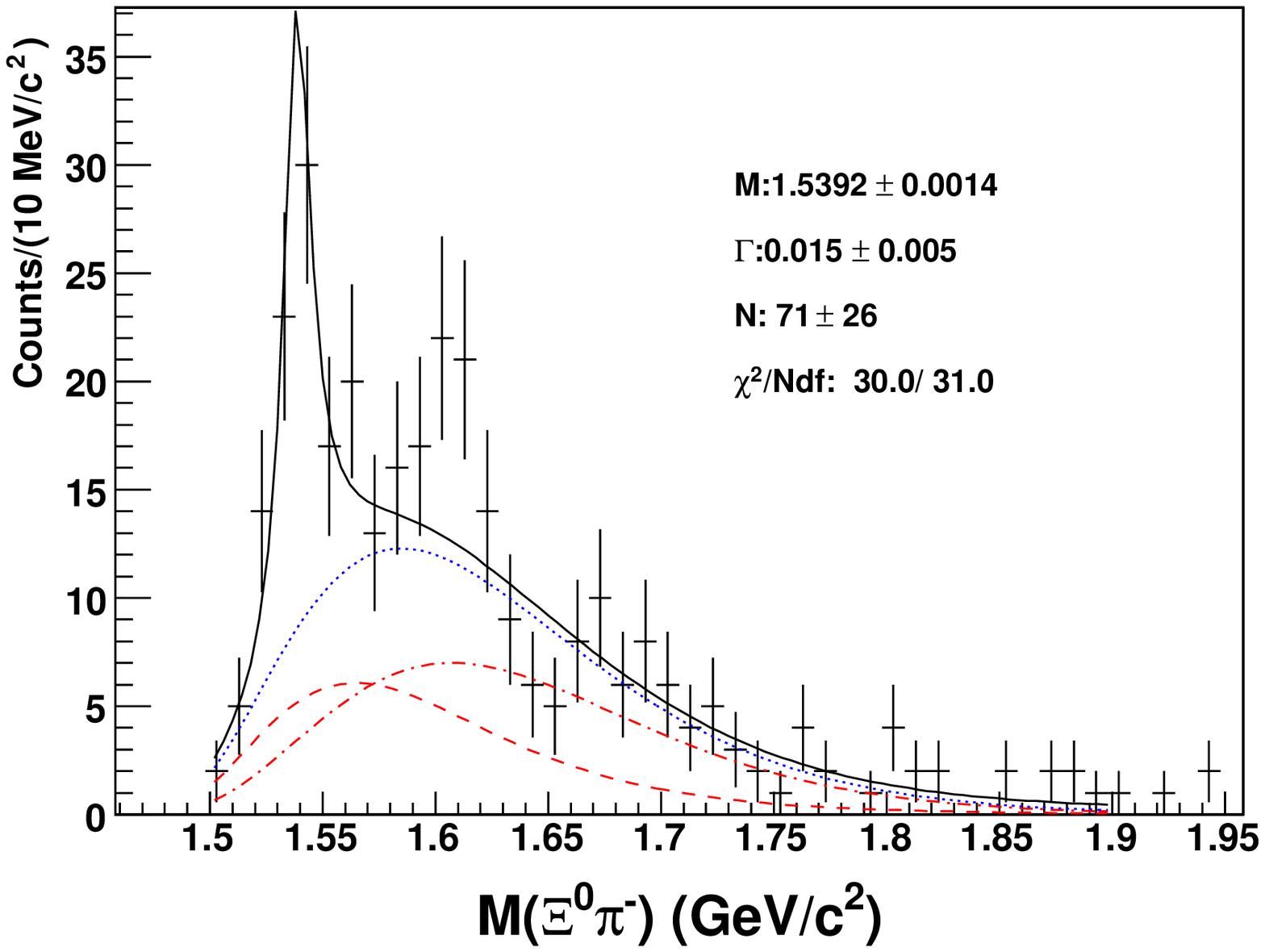}
\end{center}
\caption{Missing mass spectrum $M_X$ for $\gamma p \rightarrow K^+K^+ X$. 
The $\Xi$ ground state and first 
excited state are clearly identified. The right panel shows the invariant mass 
of $\Xi^0(1320)\pi^-$. The low mass peak is the $\Xi(1530)$. 
The other structure near 1620 MeV is statistically not significant. }
\label{fig:Xi_mm}
\label{fig:Xi_pi}
\end{figure}

The production of $\Xi$ baryons, i.e. strangeness $S = -2$ excited states, provides another 
promising way of searching for new baryon states due to the 
expected narrower widths of these states compared to $S = 0$ and $S = -1$ 
resonances. However,
the disadvantage of using a photon beam 
for detection of the $S=-2$ states is that they require production of 
at least two kaons in the final state. Possible production 
mechanisms include t-channel $K$ or $K^*$ 
exchanges on proton targets with an excited hyperon $Y^*$ ($\Lambda^*$ or $\Sigma^*$) as 
an intermediate state and 
with subsequent decays $Y^* \rightarrow K^+ + \Xi^*$ and $\Xi^* \rightarrow \Xi \pi$ or
 $\Xi \rightarrow \Lambda (\Sigma) \bar{K}$.  The missing mass technique may work for 
$\gamma p \rightarrow K^+ K^+ X$ if the state is 
sufficiently narrow to be observed as a peak in the missing mass spectrum. 
Fig. \ref{fig:Xi_mm} shows that one can 
identify the lowest two cascade states in this way. 
To identify the higher-mass states, higher energies are needed.

Another possibility to isolate excited cascade baryons is to measure 
additional particles in the 
final state resulting from the decays of the excited $\Xi$,
e.g. $K^+K^+\Xi^0\pi^-$.
The invariant mass of the $\Xi\pi^-$ system 
shows the first exited state $\Xi(1530)$ and indications of 
additional structure near 1620 MeV.  A state 
near that mass is predicted as a dynamically generated $\Xi\pi$ system 
in the model of Ref. \cite{ramos}. The 
data have insufficient statistics and were taken at too low
of an energy 
to allow further investigations. New 
data taken at 5.5 GeV electron energy and with higher statistics 
are currently being analyzed, and should 
allow more definite conclusions.  

\subsection{Demise of the pentaquarks?}

In 2003 and 2004, about a dozen experiments in nuclear physics 
and high energy physics 
laboratories claimed 
the observation of several states with flavor exotic quantum numbers, 
$\Theta^+_5(1540)$, 
$\Phi_5(1860)$, and $\Theta_c^0(3100)$. 
For a summary see Ref. \cite{burkert06}. These claims 
were exclusively 
based on non-dedicated experimental observations. 
At JLab a series of three high statistics 
dedicated measurements 
was conducted in 2004 and 2005 with CLAS to verify 
some of these claims. 
None of the earlier 
observations of a
$\Theta^+_5(1540)$ 
could be confirmed and stringent upper limits 
on production cross sections 
were placed in several channels, in particular,  
$\gamma p \rightarrow \bar{K^0} K^0$, $\bar{K^0} K^+ n$ 
\cite{devita06,battag06}, 
$\gamma n \rightarrow K^- K^+ n$ \cite{mckinnon06}, 
$\gamma D \rightarrow \Lambda n 
K^+$ \cite{nicc06}, 
$\gamma p \rightarrow K^+ K^- p$ \cite{kubar06}. 
Other channels are still being evaluated.

\section{Results from Electroproduction Experiments}
\subsection{JLab/CLAS data on single pion electroproduction}

The CLAS detector at
Jefferson Lab
is the first full acceptance instrument
designed for the comprehensive investigation
of exclusive electroproduction of mesons with the goal
to study the electroexcitation of nucleon
resonances in detail.
In recent years, a variety of measurements
of single pion electroproduction on protons have been performed
at CLAS in a wide range of $Q^2$ from 0.16 to 6 GeV$^2$
\cite{Joo1}-\cite{Biselli}.
The obtained experimental data include about
119,000 data points of differential cross sections, 
longitudinally polarized
beam asymmetries, and longitudinal target and 
beam-target asymmetries.
A comprehensive analysis of these data was performed
using two approaches: one, based on fixed-$t$ dispersion
relations (DR), and another, based on the effective Lagrangian
approach - Unitary Isobar Model (UIM).

The DR approach was developed for pion photo- and 
electroproduction on nucleons
in the 50s
\cite{Chew,Fubini} and played an extremely fruitful
role in the analyses of the data for these processes. 
In recent publications \cite{Aznauryan1998,Aznauryan2003,Aznauryan2008,
Aznauryan2009},
this approach was further developed by extending it into
the wide kinematical region in $W$ and $Q^2$
covered by the new data.
DR allowed also to obtain 
strict constraints on the multipole 
amplitudes
$M_{1+}^{3/2}$, $E_{1+}^{3/2}$, $S_{1+}^{3/2}$
that correspond to the contribution
of the prominent $\Delta$(1232)P$_{33}$ 
resonance \cite{Aznauryan2003}.
The constraint on the large
$M_{1+}^{3/2}$ amplitude plays an important role in the reliable
extraction of the amplitudes for  the
$\gamma^* N\rightarrow \Delta$(1232)P$_{33}$ transition.
It also impacts the analysis
of the second resonance region,
because resonances from this
region are overlapping with the $\Delta$(1232)P$_{33}$.

Starting in the late 90s, another approach, the 
UIM \cite{Drechsel} (also known as MAID), became widely
used for
the analyses of single-pion photo- and electroproduction
data. This approach was modified in Ref. \cite{Aznauryan2003}  by the
incorporation of Regge poles, which
enabled a good description of all photoproduction multipole
amplitudes  
up to an invariant mass
$W=2$~GeV using a unified Breit-Wigner parametrization of the
resonance contributions.

Within DR and UIM,
the analyses of the data
\cite{Joo1}-\cite{Biselli}
were performed  
in Refs. \cite{Aznauryan2008,
Aznauryan2009,
Aznauryan2005,Aznauryan2005_1}. 
As a result, the helicity amplitudes for the 
electroexcitation
of the low mass resonances
$\Delta$(1232)P$_{33}$, N(1440)P$_{11}$,
N(1520)D$_{13}$, and
N(1535)S$_{11}$ 
were extracted from the
experimental data in a range of  
$Q^2$ up to 6 GeV$^2$.   
The non-resonant contributions are built in DR and UIM
in conceptually different ways;
this allowed to draw conclusions on the model
sensitivity of the  resulting electroexcitation amplitudes.
Model uncertainties caused by the higher-mass
resonances and by the non-resonant amplitudes
due to the uncertainties in the Born terms (the $s$- and $u$-channel 
nucleon exchanges
and $t$-channel pion contribution) and in the $\rho$ and $\omega$ 
$t$-channel contributions were also evaluated.
All uncertainties 
were added in quadrature and presented as
model uncertainties of the extracted electroexcitation
amplitudes.

\begin{figure}
\begin{center}
  \includegraphics[height=.28\textheight]{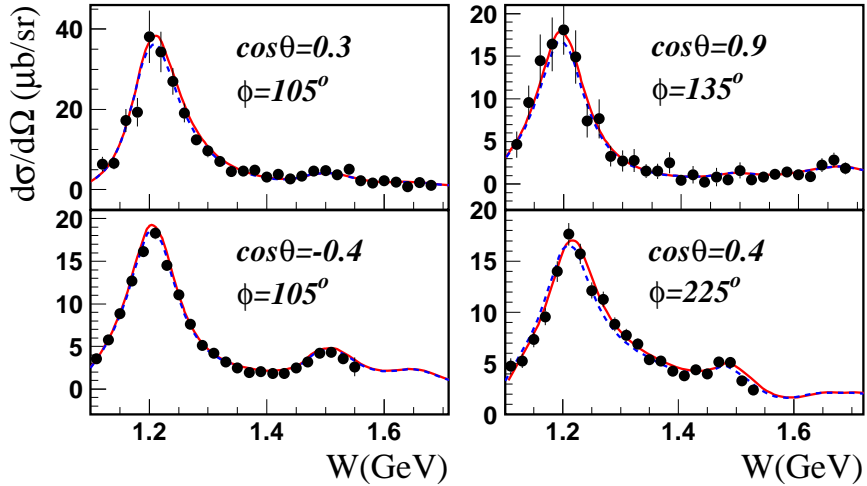}
\end{center}
  \caption{
$W$-dependence of the differential cross
sections for $\gamma^* p\rightarrow \pi^0 p$ (upper row)
and $\gamma^* p\rightarrow \pi^+ n$ (lower row)
at $Q^{2}=0.4~$GeV$^{2}$ for different polar and azimuthal angles.
Data are from Refs. \cite{Joo1} ($\pi^0$)
and \cite{Egiyan} ($\pi^+$).
The solid (dashed) curves
correspond to the results obtained using the DR (UIM) approach
\cite{Aznauryan2009}.
}  
\label{fig:w_04}
\end{figure}

\begin{figure}
\begin{center}
  \includegraphics[height=.4\textheight]{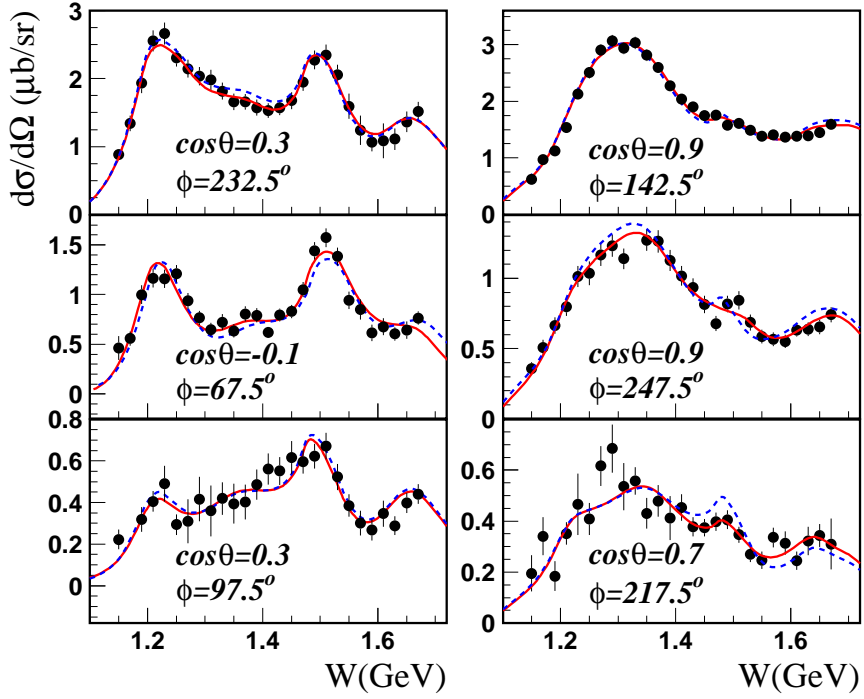}
\end{center}
 \caption{
$W$-dependence of the differential cross
section for $\gamma^* p\rightarrow \pi^+ n$
for different polar and azimuthal angles.
Upper, middle, and lower rows are for
$Q^{2}$=1.72,~2.44,~3.48 GeV$^{2}$, respectively.
Data are from Ref. \cite{Park}.
Other notations as in Fig. \ref{fig:w_04}.
}
\label{fig:w_high}
\end{figure}

\begin{figure}
\begin{center}
  \includegraphics[height=0.27\textheight]{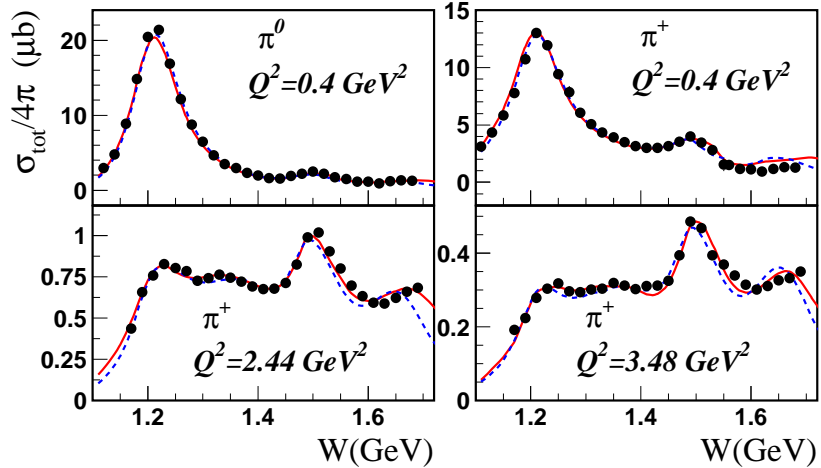}
\end{center}
  \caption{
$W$-dependence of the
$\gamma^* p\rightarrow \pi^0 p,\pi^+ n$
total cross sections.
Data are from Refs. \cite{Joo1,Egiyan,Park}.
Other notations as in Fig. \ref{fig:w_04}.
}
\label{fig:sigma}
\end{figure}

\begin{figure}
\begin{center}
  \includegraphics[height=.27\textheight]{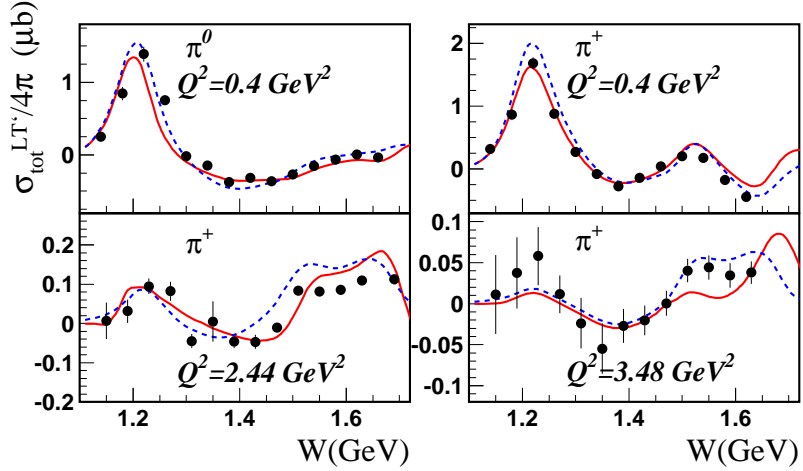}
\end{center}
  \caption{$W$-dependence of the structure
function $\sigma_{tot}^{LT'}$ corresponding to a longitudinally
polarized electron beam integrated over polar angle.
Data are from Refs.~\cite{Joo2,Joo3,Park}.
Other notations as in Fig.~\ref{fig:w_04}.}
\label{fig:asym}
\end{figure}

\begin{figure}
\begin{center}
  \includegraphics[height=.27\textheight]{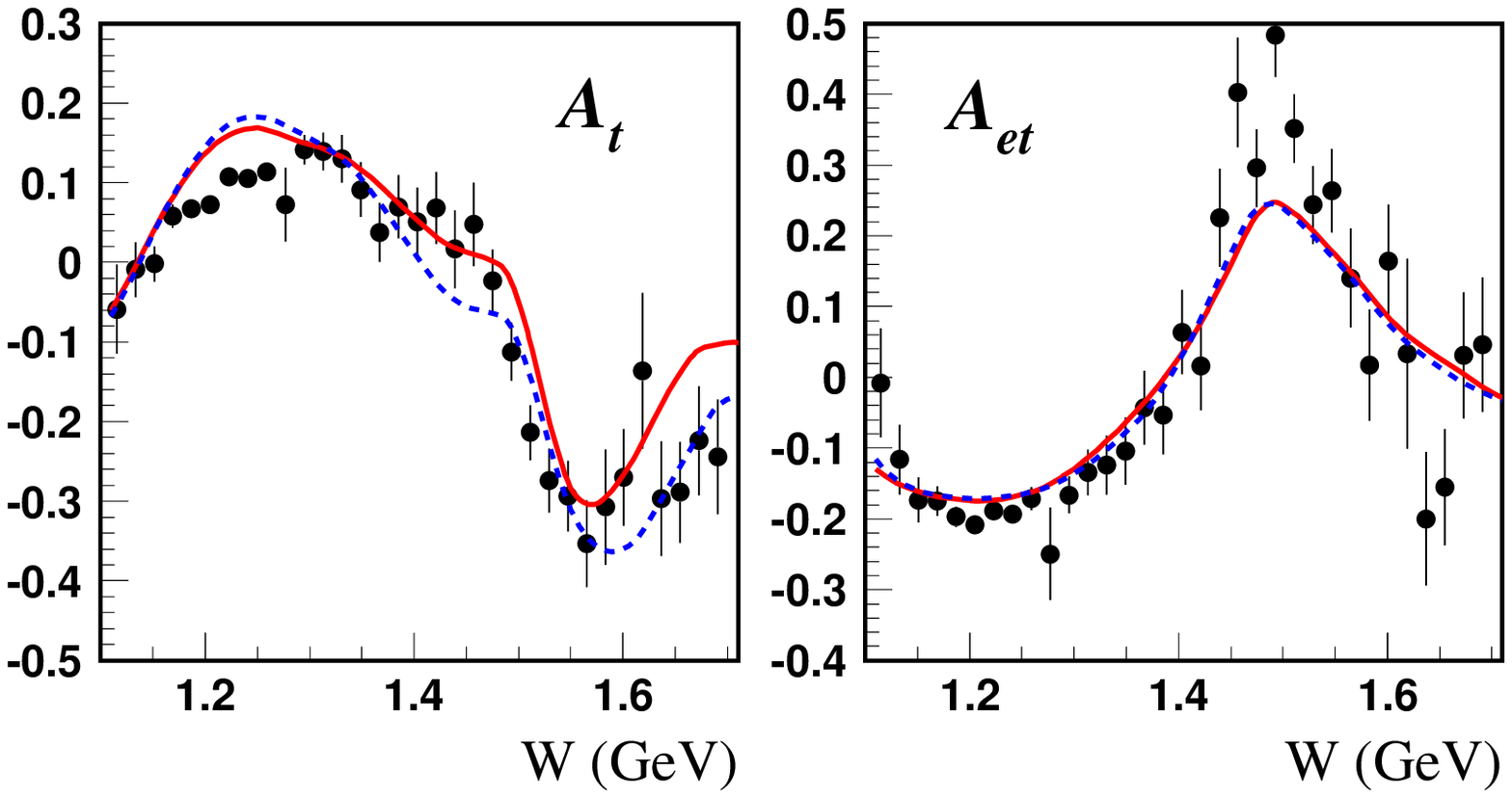}
\end{center}
  \caption{
$W$-dependence of the
target ($A_{t}$) and beam-target ($A_{et}$) asymmetries
for $\vec{e}\vec{p}\rightarrow ep\pi^0$
integrated over  cos$\theta$, $\phi$ and $Q^2$
($0.252<Q^2<0.611~$GeV$^2$).
Data are from Ref. \cite{Biselli}.
Other notations as in Fig. 
\ref{fig:w_04}.
}
\label{fig:at_aet}
\end{figure}

In Figs. \ref{fig:w_04}-\ref{fig:at_aet}, some examples that show
the description of the data are presented.
Shown are the $W$-dependences of the differential cross sections
for different reactions at several $Q^2$, polar and
azimuthal angles. The $W$-dependences of $\sigma_{tot}$, 
$\sigma^{LT'}_{tot}$, and of the
target ($A_{t}$) and beam-target ($A_{et}$) asymmetries
for $\vec{e}\vec{p}\rightarrow ep\pi^0$
integrated over  cos$\theta$, $\phi$ and $Q^2$
are also shown.


\subsection{JLab/CLAS data on two pion electroproduction}

Two pion electroproduction is 
one of the biggest contributors 
to the $N^*$ electroexcitation. 
This exclusive
channel is sensitive to almost all well established 
excited  states and in
particular to high lying resonances with masses above $1.6~$GeV. 
According to the
quark model expectations \cite{Ca00,Ca94}, 
two pion electroproduction has a 
great potential for the
discovery of the so-called "missing" baryon states. 

A combination of the CEBAF continuous electron beam and 
detector CLAS 
for the first time makes it possible to measure nine independent 
one-fold differential
$\gamma^* p\rightarrow \pi^+\pi^-p$ cross sections,
as well fully integrated cross sections, in the 
kinematical areas
presented in Table~\ref{2picover}. 
The high statistics and good momentum
resolutions of the measurements allowed  
to use bin sizes of $W$ and $Q^2$,
which are several times smaller than 
the ones used in previous measurements.
    
\begin{table}
\begin{center}
\begin{tabular}{|c|c|c|c|}
\hline
$Q^{2}$ coverage , &  $W $coverage, & Bin size over $W/Q^2$ , & Data status \\
GeV$^2$  &  GeV  &   GeV/GeV$^2$  &   \\
\hline
0.20-0.60 & 1.30-1.57 & 0.025/0.050  & Completed \cite{Fe09}  \\
0.50-1.50 & 1.40-2.10 & 0.025/0.3-0.4 & Completed \cite{Ri03} \\
2.0-5.0 & 1.40-2.00 & 0.025/0.5 & In progress  \\
  0.   & 1.60-2.80 &  0.025  & In progress  \\
\hline
\end{tabular}
\caption{\label{2picover}Kinematical areas covered by the CLAS 
measurements of the $\pi^+\pi^-p$
photo- and electroproduction cross sections.}
\end{center}
\end{table}

All essential mechanisms contributing to $\pi^+\pi^-p$ 
electroproduction were established from
the analysis of these data within
the framework of the reaction model presented
in Ref. \cite{Mo09}.
This model provides good
data description, allowing to isolate resonant contributions to observables,
needed for evaluation of  $\gamma^*NN^*$ electrocouplings.

\begin{figure}
\begin{center}
  \includegraphics[height=.5\textheight]{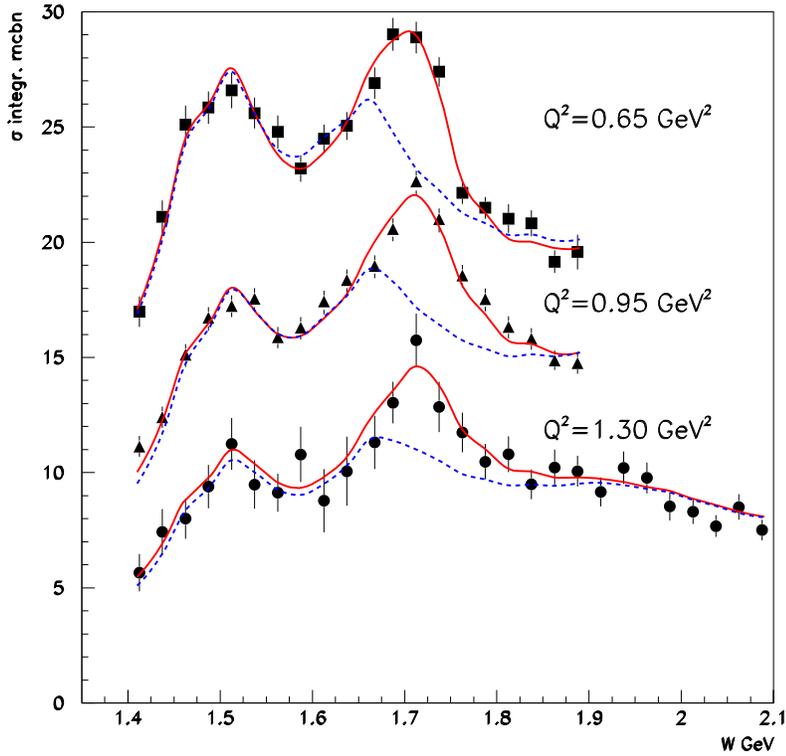}
\end{center}
\caption{
Description of the CLAS  data 
\cite{Ri03} on fully integrated 
$\pi^+\pi^-p$ electroproduction cross sections within the framework of
the reaction model \cite{Mo09} with parameters fitted to one-fold differential 
cross sections. 
The calculations, 
taking into account the contributions from the conventional $N^*$'s only, 
are shown by the dashed lines,
while the solid lines correspond to the fit after 
implementation of the $3/2^+(1720)$ 
candidate state. Difference between solid and dashed lines 
represents a signal from the 
possible new state.} 
\label{fig:missing}
\end{figure}

The CLAS data \cite{Ri03} for the first time revealed the structure 
in the fully integrated $\pi^+\pi^-p$
electroproduction cross sections  at $W$ $\approx $ 1.7 $GeV$,
shown in Fig.~\ref{fig:missing}.
Analysis of these data  within the framework of 
the reaction model \cite{Mo09} 
offers an evidence for the
candidate state $3/2^+(1720)$. Signals from
this possible new state 
can be seen in Fig.~\ref{fig:missing}. If these signals will be
confirmed in the future combined analysis of the  CLAS $\pi^+\pi^-p$
photo- and electroproduction data (Table~\ref{2picover}), we 
would obtain an unambiguous  evidence for long awaited "missing" baryon state. 

Evaluation of  $\gamma^*NN^*$ electrocouplings from the CLAS  
$\pi^+\pi^-p$ electroproduction data is in progress.
The $\gamma^*NN^*$ electrocouplings for almost all $N^*$ states 
will be determined at photon
virtualities  $Q^2<5.0~$GeV$^2$. Preliminary results
may be found in Refs. \cite{Mo091,Mo10}. Single and charge double pion
electroproduction channels have completely different non-resonant contributions
and provide independent information on the extracted resonance electrocouplings.
The consistent results obtained from analyses of the dominant $\pi N$ and
$\pi^+\pi^-p$ exclusive channels will offer an evidence for reliable 
extraction of these
fundamental quantities from the meson electroproduction data.

Finally, we are planning a combined analysis of single and charge double pion
electroproduction data within the framework of advanced 
coupled-channel approach, that
is described in Sec. 4.

\subsection{Other JLab data on exclusive meson electroproduction}
There are JLab Hall A and Hall C measurements
of $\pi^0$ electroproduction on the proton
in the $\Delta$(1232)P$_{33}$ resonance region that 
include Hall A data on recoil polarization response 
functions at $Q^2=1~$GeV$^2$ \cite{KELLY1,KELLY2} 
and Hall C data on differential cross sections at $Q^2=2.8,~4.2~$GeV$^2$ 
\cite{Frolov} and 
$6.4,~7.7~$GeV$^2$ 
\cite{Vilano}.
Measurements of $\eta$ electroproduction 
on the proton were performed in Hall B with  CLAS and in Hall C;
they include 
CLAS data at 
$Q^2=0.165-3.1~$GeV$^2$ \cite{Thompson,Denizli} and 
Hall C data at $Q^2=2.4,~3.6~$GeV$^2$ \cite{Armstrong},
$Q^2=5.7,~7~$GeV$^2$ \cite{Dalton}.

\subsection{The $\Delta$(1232)P$_{33}$ resonance}
\begin{figure}
  \includegraphics[height=.3\textheight]{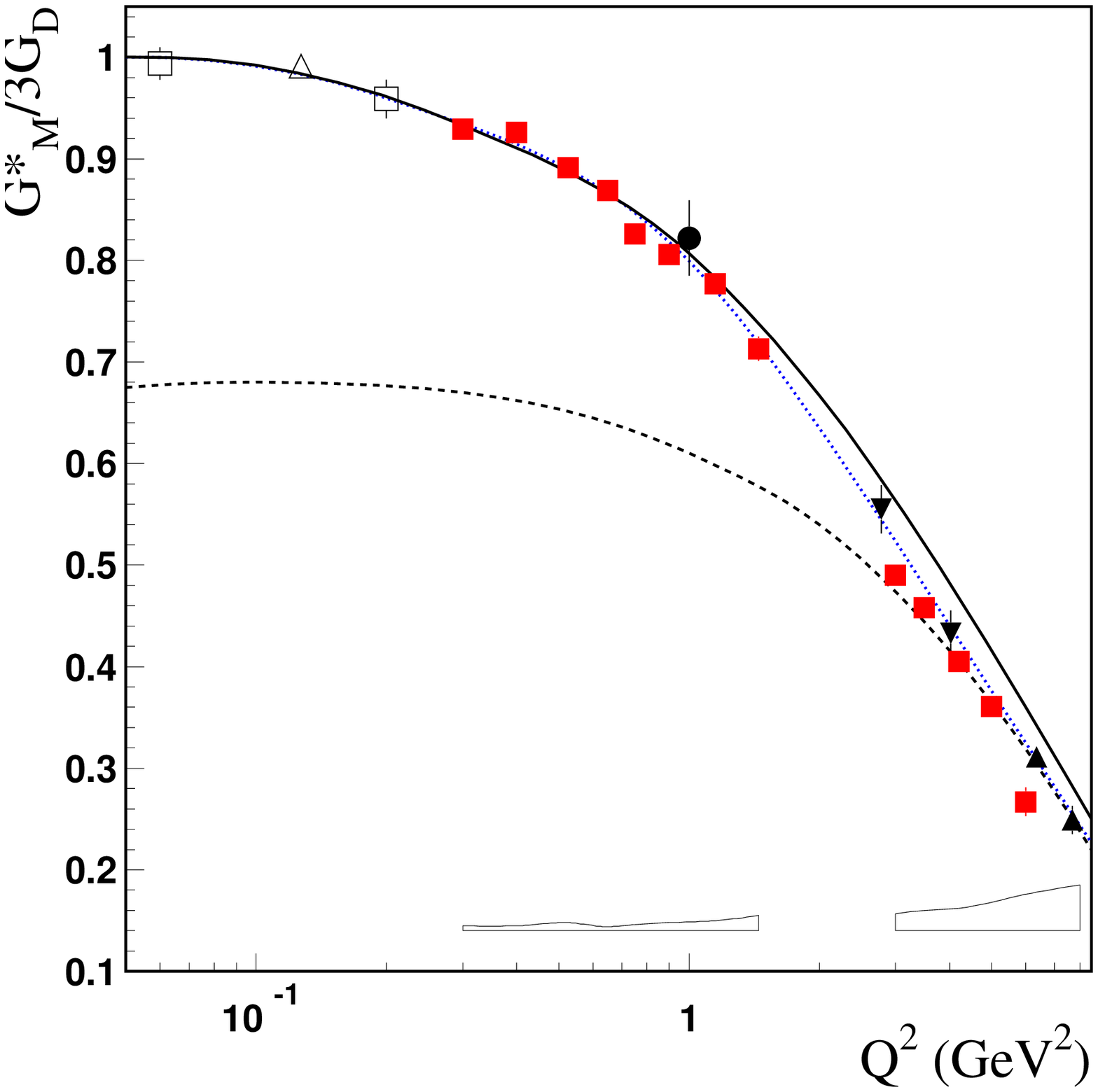}
  \includegraphics[height=.3\textheight]{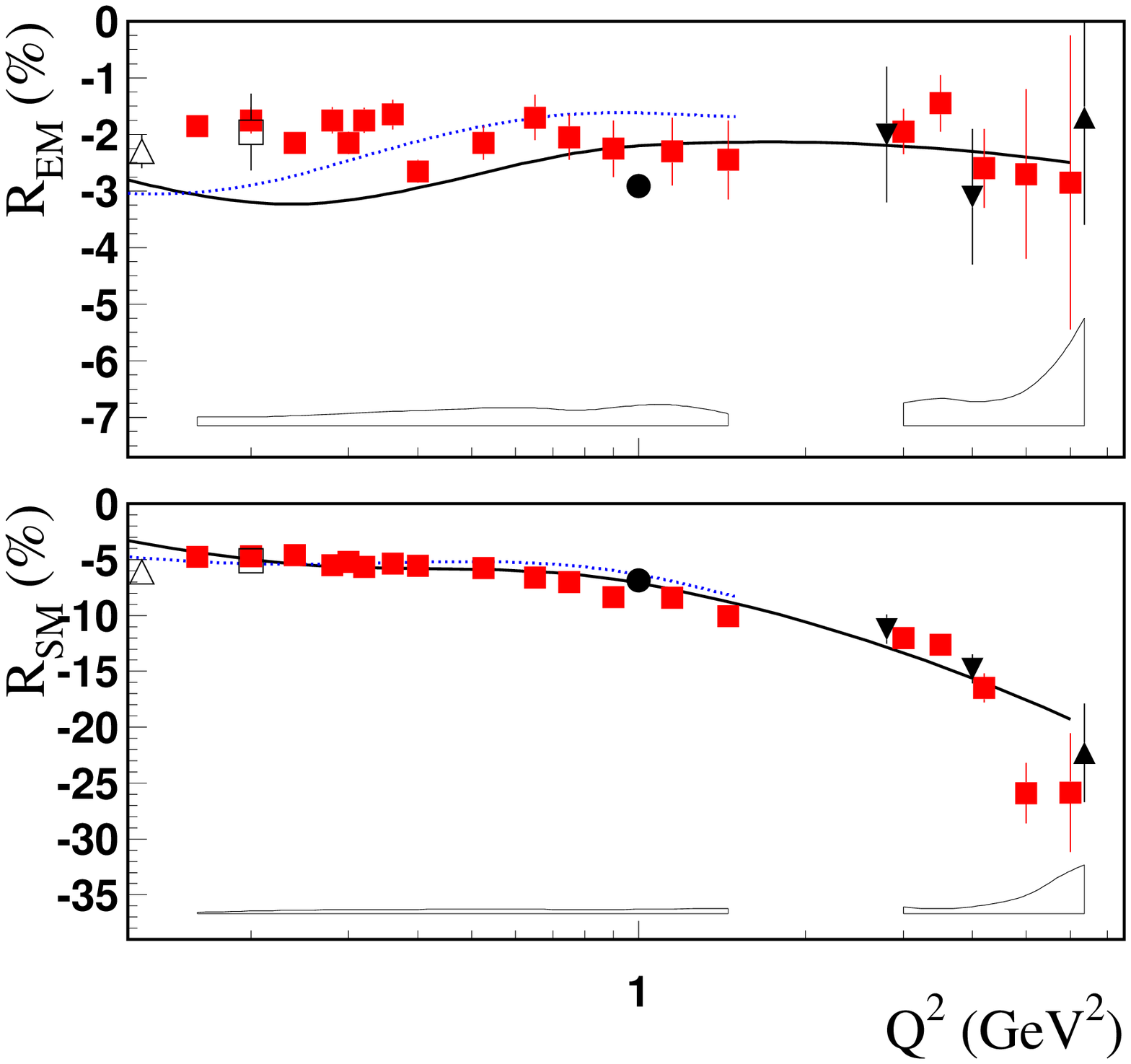}
  \caption{
Left panel: the form factor $G^*_{M}$  
\cite{Ash}
for the  $\gamma^* p \rightarrow~\Delta$(1232)P$_{33}$
transition relative to $3G_D(Q^2)$: 
$G_D(Q^2)=1/(1+\frac{Q^2}{0.71~GeV^2})$.
Right panel: the ratios $R_{EM}\equiv E_{1+}^{3/2}/M_{1+}^{3/2}$,
$R_{SM}\equiv S_{1+}^{3/2}/M_{1+}^{3/2}$.
The solid boxes are
the results extracted
from the CLAS data \cite{Joo1,Joo2,Ungaro,Smith}
in Ref. \cite{Aznauryan2009}.
The bands show the model uncertainties of the results.
Also shown are the results from
JLab/Hall A \cite {KELLY1,KELLY2} - solid circles
and JLab/Hall C \cite{Frolov,Vilano} - solid triangles.
The low $Q^2$ results are from
MAMI \cite{MAMI006,MAMI02}  - open boxes and
MIT/BATES \cite{BATES} - open triangles.
The solid and dashed curves correspond to
the `dressed' and `bare' contributions from Ref. \cite{Sato};
for $R_{EM},~R_{SM}$, only the `dressed'
contributions are shown; the
`bare' contributions are close to zero.
The dotted curves are the predictions obtained
in the large-$N_c$ limit of QCD \cite{GPD1,Pascalutsa}.
}
\label{fig:delta}
\end{figure}

The excitation of the $\Delta$(1232)P$_{33}$  has been studied for more
than 50 years with various probes. But only in the past decade
have the experimental tools become available in electron
scattering to enable precise measurements
in exclusive $\pi$ production from protons with
photon virtualities up to $Q^2=8~$GeV$^2$.
Benchmark results from CLAS 
and other laboratories
are shown in Fig. \ref{fig:delta}.
New exclusive measurements confirm a
rapid falloff of the $\gamma^* p \rightarrow~\Delta$(1232)P$_{33}$
magnetic-dipole form factor relative to that for the proton
seen previously in 
inclusive experiments.
There has been a long-standing underestimation of the data for this form factor 
by the constituent quark model.
Within
dynamical reaction models \cite{Sato,Kamalov,msl07}, 
the pion-cloud contribution
was identified as
the source of this discrepancy.
The importance of the pion-cloud contribution
for the $\gamma^* p \rightarrow~\Delta$(1232)P$_{33}$
transition
is confirmed also by the lattice QCD calculations \cite{Alexandrou}.
In Fig. \ref{fig:delta}, the results of the
dynamical model \cite{Sato} are given that show
that the pion-cloud contribution
makes up more than 30\% of $G_M^*(Q^2)$
at the photon point, and remains sizeable at the highest $Q^2$.

The $Q^2$-dependence
of  $R_{EM}$ and $R_{SM}$ is of great interest
as a measure of the $Q^2$ scale where the approach
to the asymptotic domain
of QCD may set in.
In the pQCD asymptotics $R_{EM}\rightarrow 1$ and
$R_{SM}\rightarrow const$. The data
on $R_{EM},R_{SM}$ show that in the range
$Q^2< 7~$GeV$^2$, there is no sign of an approach to the
asymptotic pQCD regime in either of these ratios.

\subsection{The Roper resonance N(1440)P$_{11}$ - a puzzle resolved}
\begin{figure}
\begin{center}
  \includegraphics[height=.23\textheight]{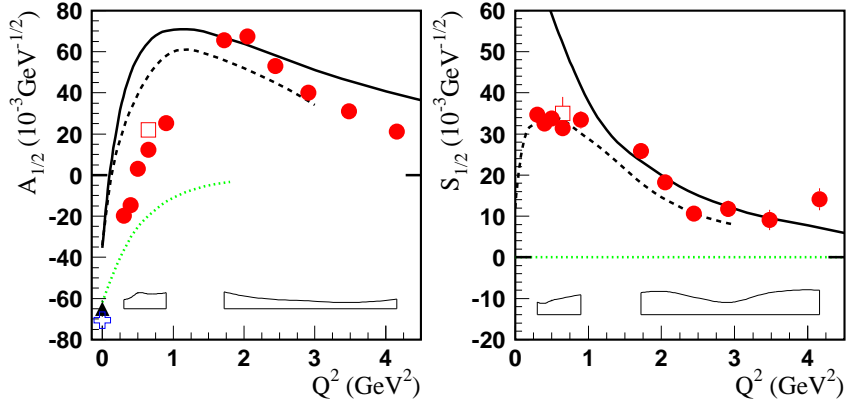}
\end{center}
  \caption{
Helicity amplitudes
for the  $\gamma^* p \rightarrow~$N(1440)P$_{11}$
transition.
The solid circles are
the results extracted from
the CLAS data \cite{Joo1,Joo2,Joo3,Egiyan,Park,Biselli}
in Ref. \cite{Aznauryan2009}.
The bands show the model uncertainties.
The cross at $Q^2=0$ is from SAID analysis
that includes CLAS data on $\pi^0$
and $\pi^+$ photoproduction cross sections \cite{dugger07,dugger09}.
The open boxes are obtained in the combined analysis
of CLAS single $\pi$ and 2$\pi$ electroproduction data
\cite{Aznauryan2005_1}.
The full triangle at $Q^2=0$ is the RPP estimate \cite{pdg2008}.
All amplitudes, except RPP values, correspond to
$M=1440~$MeV, $\Gamma_{tot}=350~$MeV,
and  $\beta_{\pi N}=0.6$.
The predictions of the
LF relativistic quark models 
\cite{Capstick,AznRoper} 
assuming N(1440)P$_{11}$
is a first radial excitation of the $3q$ ground state are shown
by solid and dashed curves.
The dotted curves are
obtained assuming that the N(1440)P$_{11}$
is a gluonic baryon excitation (q$^3$G hybrid state) \cite{Li1,Li2}.
}
\label{fig:p11}
\end{figure}

By quantum numbers, the simplest and most natural classification of the
Roper resonance in the constituent quark model is as a first
radial excitation of the $3q$ ground state. However, difficulties
of quark models to describe the low mass and large
width of the N(1440)P$_{11}$,
and also its electroexcitation amplitudes
on proton and neutron at the photon point, gave rise to numerous
speculations around this state. Alternative descriptions
of this state as a
gluonic baryon excitation \cite{Li1,Li2}, or as a hadronic 
N$\sigma$ molecule \cite{Krehl}, were suggested.
The CLAS measurements, for the first time, allowed the 
determination of the electroexcitation
amplitudes of the Roper resonance on the proton 
in the range $Q^2<4.5$ GeV$^2$ 
(Fig. \ref{fig:p11}). 
These results are crucial for
the understanding of the nature of this state.
There are several specific features in the extracted
$\gamma^* p \rightarrow~$N(1440)P$_{11}$ amplitudes
that are very important for testing models.
First, this is the specific behavior of the transverse amplitude $A_{1/2}$,
which being large and negative at $Q^2=0$,
becomes large and positive at $Q^2\simeq 2~$GeV$^2$, and
then drops smoothly with $Q^2$. Second, 
the relative sign between the longitudinal $S_{1/2}$ and
transverse $A_{1/2}$ amplitudes. 
And third, 
the common sign of the amplitudes $A_{1/2},S_{1/2}$ extracted from the data
on $\gamma^*p\rightarrow \pi N$ includes signs 
from the $\gamma^*p\rightarrow$N(1440)P$_{11}$ 
and N(1440)P$_{11}\rightarrow \pi$N vertices; 
both signs should be taken
into account while comparing with model predictions.
All these characteristics are described by the 
light-front relativistic quark models \cite{Capstick,AznRoper} assuming that 
N(1440)P$_{11}$ is the first radial excitation of the $3q$ ground state.
The presentation of the Roper resonance as a 
gluonic baryon excitation is definitely ruled out. 
No predictions exist for the N$\sigma$ molecule model.

\subsection{The N(1535)S$_{11}$ state}
\begin{figure}
\begin{center}
  \includegraphics[height=.23\textheight]{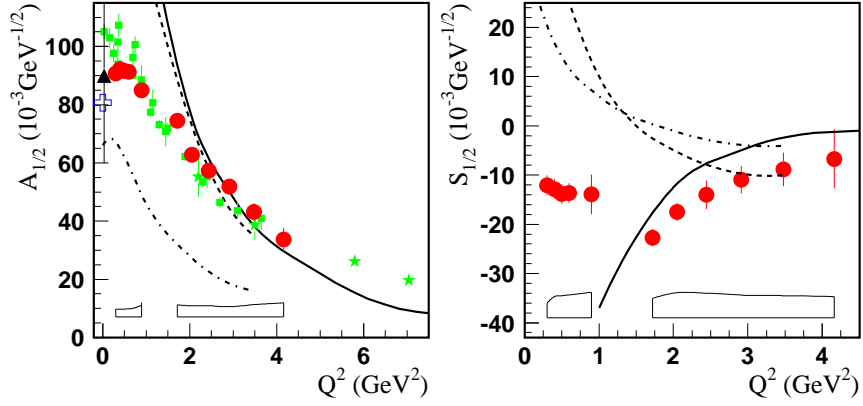}
\end{center}
  \caption{
Helicity amplitudes
for the  $\gamma^* p \rightarrow~$N(1535)S$_{11}$
transition. The legend is partly as for Fig. \ref{fig:p11}.
The solid boxes and stars are  the CLAS 
\cite{Thompson,Denizli}
and Hall C 
\cite{Armstrong,Dalton}
results, respectively, obtained
in $\eta$  electroproduction. 
All amplitudes, except RPP values, correspond to
$M=1535~$MeV, $\Gamma_{tot}=150~$MeV,
and to the branching ratios  $\beta_{\pi N}=0.485$,
$\beta_{\eta N}=0.46$ (2,3).
The results of the
LF relativistic quark models are given by the dashed  \cite{Capstick}
and dashed-dotted \cite{Simula1} curves. The solid curves
show the amplitudes found within
light-cone sum rules using lattice results for the light-cone
distribution amplitudes of the N(1535)S$_{11}$ resonance \cite{Braun}.
}
\label{fig:s11}
\end{figure}

For the first time the $\gamma^*N\rightarrow$
N(1535)S$_{11}$ transverse helicity amplitude
has been extracted from the $\pi$
electroproduction data
in a wide range of $Q^2$ (Fig. 
\ref{fig:s11}),
and the results confirm the
$Q^2$-dependence of this amplitude 
observed in $\eta$ electroproduction.
Comparison of the results
extracted from $\pi$ and
$\eta$ photo- and electroproduction data
have allowed to specify the relation between the branching ratios
of N(1535)S$_{11}$ to the $\pi N$ and $\eta 
N$ channels: from the fit to the amplitudes
at $0\leq Q^2<4.5~$GeV$^2$,
we found
\begin{equation}
\frac{\beta_{\eta N}}{ \beta_{\pi N}}=0.95\pm 0.03.
\end{equation}
Further, taking into account the branching
ratio to the $\pi\pi N$ channel
$\beta_{\pi\pi N}=0.01-0.1$ \cite{pdg2008},
which accounts
practically for all channels different
from $\pi N$ and $\eta N$,
we found
\begin{eqnarray}
&&\beta_{\pi N}=0.485\pm 0.008\pm 0.023, \\
&&\beta_{\eta N}=0.460\pm 0.008\pm 0.022.
\end{eqnarray}
The first error corresponds to the fit error in Eq. (1)
and the second error is related to the uncertainty of
$\beta_{\pi\pi N}$.
The obtained branching ratios (2,3) will  
enter into the
2010 edition of the RPP. 

Due to the CLAS measurements of $\pi$ 
electroproduction, for the first time 
the $\gamma^*N\rightarrow$
N(1535)S$_{11}$ longitudinal helicity amplitude
was extracted from experimental data.
These results are crucial for testing theoretical
models. It turned out that 
at $Q^2<3~$GeV$^2$, the sign of $S_{1/2}$ is not described by the quark
models. Combined with the difficulties of quark models to 
describe
the substantial
coupling of N(1535)S$_{11}$ to the $\eta N$ channel \cite{pdg2008}
and to strange particles \cite{Liu,Xie},
this can be indicative of 
a large meson-cloud contribution
and(or) additional
$q\bar{q}$ components
in this state \cite{An}.
Alternative representations of the N(1535)S$_{11}$
as a meson-baryon molecule have been also discussed 
\cite{Weise,Nieves,Oset1,Lutz}.

\subsection{The N(1520)D$_{13}$ resonance}

\begin{figure}
\begin{center}
  \includegraphics[height=.22\textheight]{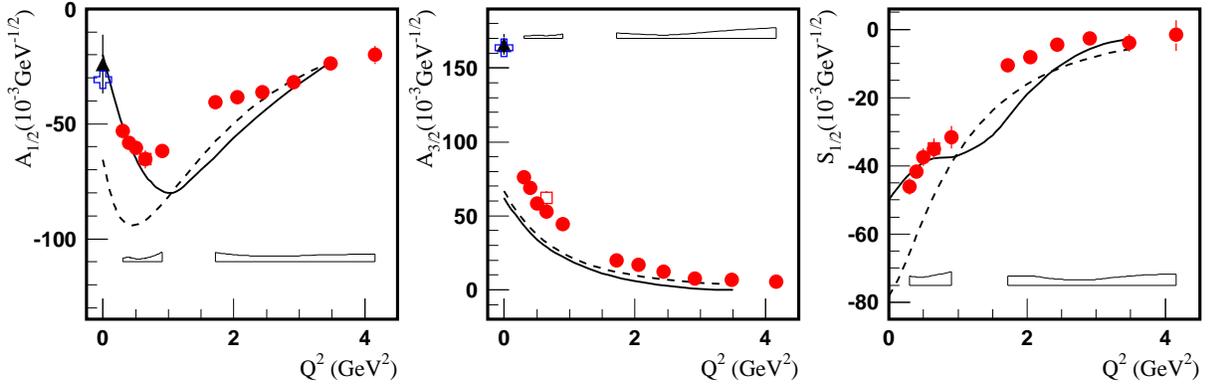}
\end{center}
  \caption{
Helicity amplitudes
for the  $\gamma^* p \rightarrow~$N(1520)D$_{13}$
transition. The legend is partly as for Fig. \ref{fig:p11}.
All amplitudes, except RPP values, correspond to
$M=1520~$MeV, $\Gamma_{tot}=112~$MeV,
and  $\beta_{\pi N}=0.6$.
The curves correspond to the predictions of
quark models:
\cite{Warns} (solid),
\cite{Santopinto} (dashed).
}
\label{fig:d13}
\end{figure}

The CLAS data allowed the determination of 
the longitudinal amplitude of
the $\gamma^* p \rightarrow ~$N(1520)D$_{13}$ 
transition, and much more
accurate results are obtained for the transverse amplitudes
(Fig. \ref{fig:d13}). 
There is a longstanding prediction of constituent
quark models for the $\gamma^* p \rightarrow~$N(1520)D$_{13}$
transition that follows from the structure 
of the nucleon and the N(1520)D$_{13}$ wave functions.
It consists in the rapid helicity switch from the 
dominance
of the $A_{3/2}$ amplitude at the photon point to the dominance 
of $A_{1/2}$ at $Q^2>1~$GeV$^2$. This prediction is definitely confirmed
by the data, although quark models fail to describe the details
of the $Q^2$ dependence of the amplitudes. 
\subsection{Transverse charge densities in the nucleon -
resonance transitions}
\begin{figure}
\begin{center}
  \includegraphics[height=.22\textheight]{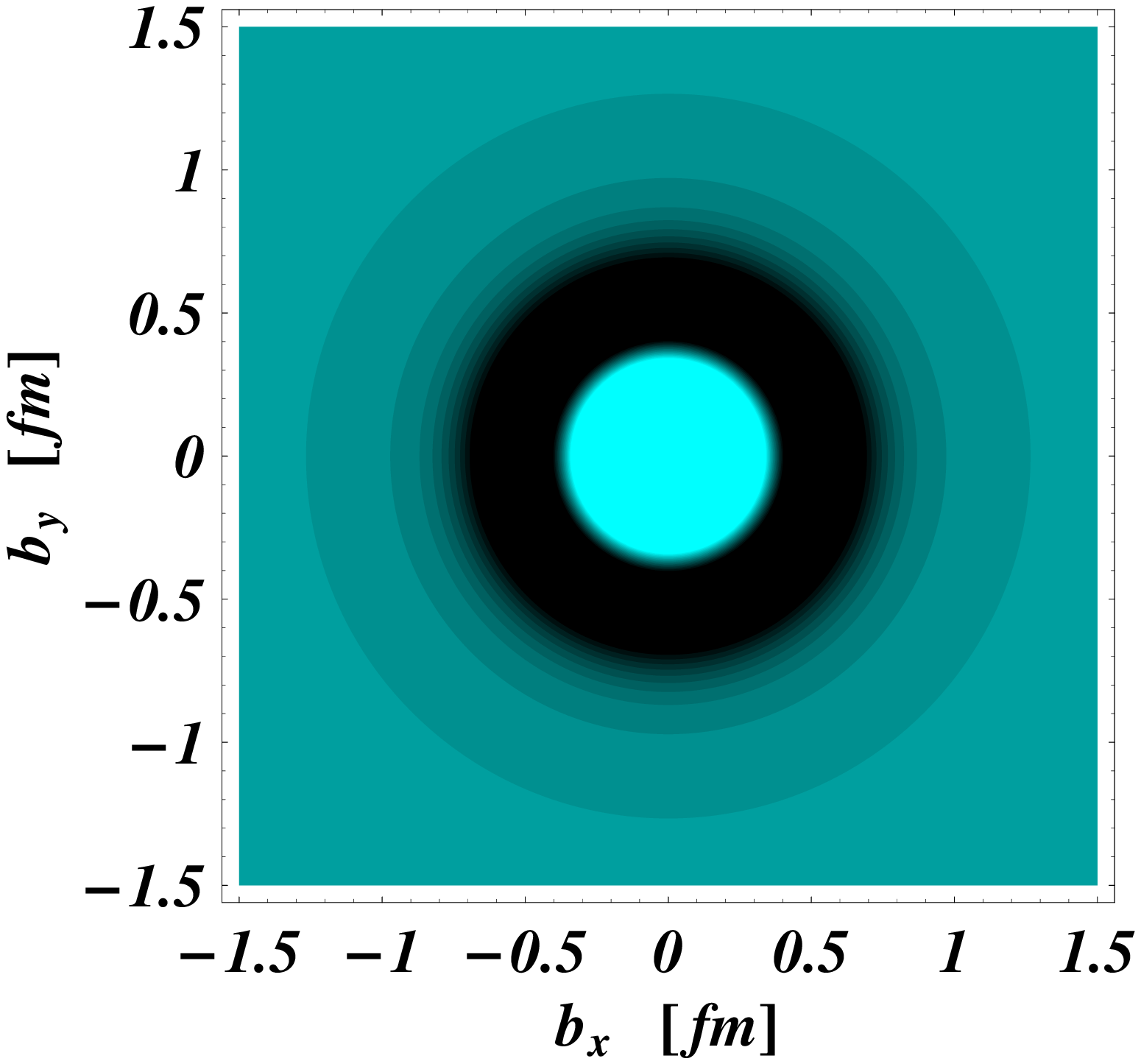}
  \includegraphics[height=.22\textheight]{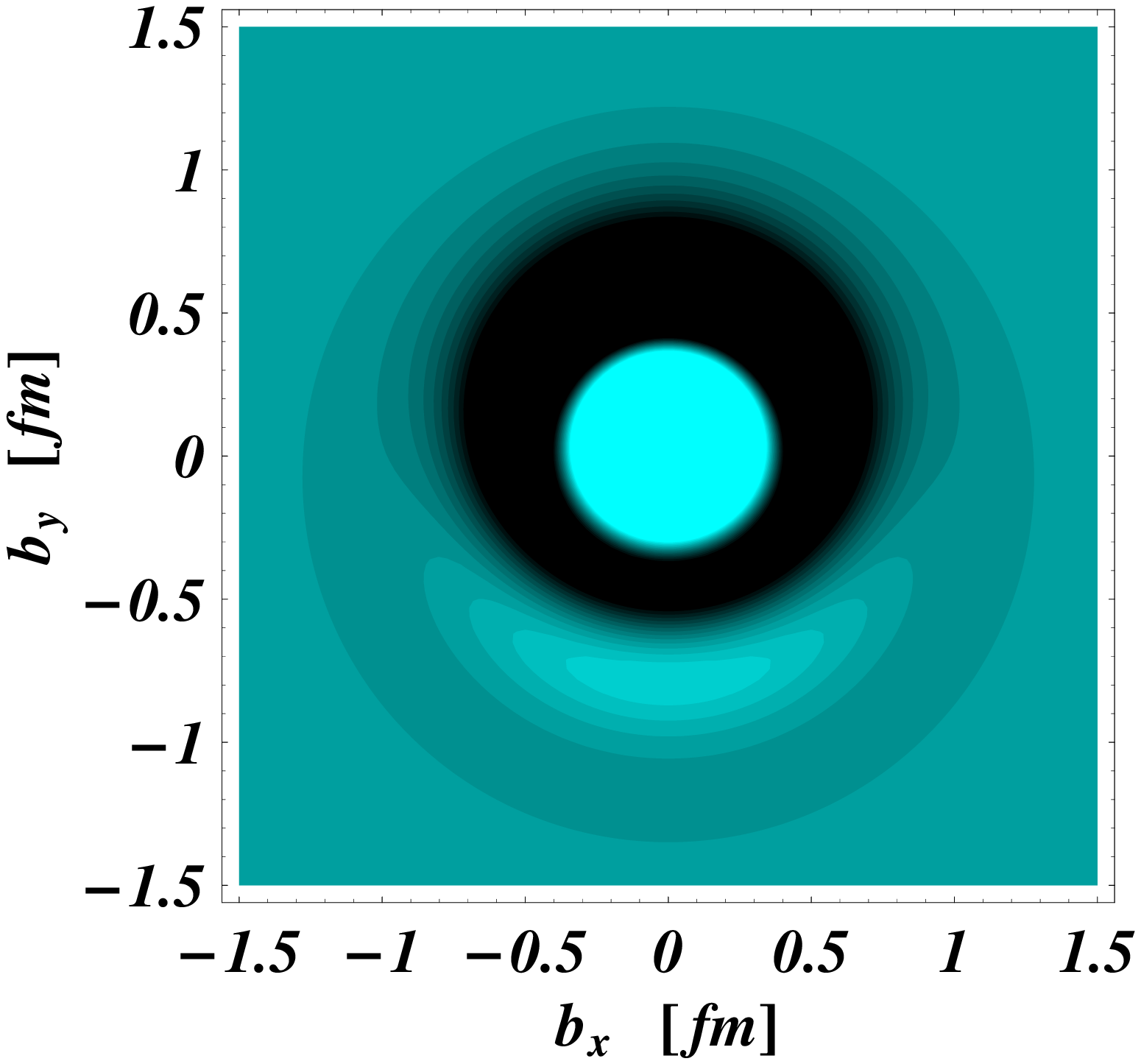}
  \caption{
Quark transverse charge density in the 
$\gamma^* p \rightarrow~$N(1440)P$_{11}$ transition 
\cite{Vanderhaeghen_den}.
Left panel: unpolarized $p$ and N(1440)P$_{11}$. Right panel:
$p$ and N(1440)P$_{11}$ polarized along the positive $x$-axis.
The light (dark) areas are dominated by up (down) quarks
and correspond to dominantly positive (negative) charges.
}
\label{fig:p11_den}
\end{center}
\end{figure}

The accurate information 
now available for all amplitudes of the
$\gamma^*p\rightarrow\Delta$(1232)P$_{33}$, N(1440)P$_{11}$,
N(1520)D$_{13}$, and
N(1535)S$_{11}$ transitions   allowed, for the first time, to obtain
spatial images of the transitions by mapping out
the transition charge densities 
in impact parameter space.
The corresponding result for the Roper resonance 
is presented in Fig. \ref{fig:p11_den}.
 
\section{Excited Baryon Analysis Center}

The Excited Baryon Analysis Center (EBAC) was established at JLab
in January, 2006 to provide theoretical support to the excited baryon program.
EBAC's program has two objectives. First, to establish the spectrum of
excited nucleon states ($N^*$) and to
extract the $N^*$ parameters from the $\pi N$, $\gamma N$ and $N(e,e'\pi)$ data.
The second objective is
to develop theoretical interpretations of the extracted $N^*$ parameters.
To achieve these two goals, a dynamical coupled-channel
reaction model (DCC) has been developed and is being used to perform
analyses of the meson production data from JLab and other facilities.
The essential feature of the DCC model is to account for the coupled-channel 
effects resulting
from the unitarity conditions and the reaction mechanisms in the
short range (off-shell) region where we want to map out the structure
of baryons. 

\begin{figure}[t]
\vspace{10pt}
\begin{center}
\mbox{\epsfig{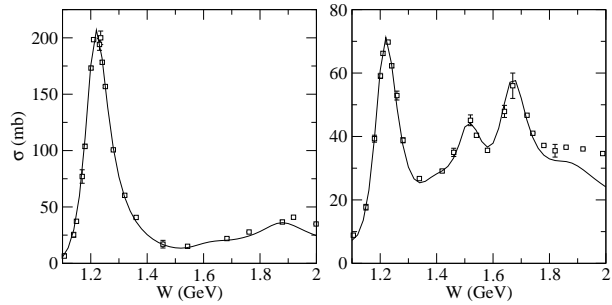}}
\end{center}
\caption{Total cross sections of $\pi^+ p$ (left) and $\pi^- p$ (right) reactions.
The curves are from the dynamical coupled-channel model
\cite{jlms07,kjlms09}.  }
\label{fig:pin}
\end{figure}

The DCC model is based on a Hamiltonian formulation \cite{msl07} of
multi-channel and multi-resonance reactions. 
The first work at EBAC was to determine the hadronic parameters of
the model by fitting the world data on the $\pi N\rightarrow \pi N$,
$\pi N \rightarrow \eta N$, and
$\pi N \rightarrow \pi \pi N$ reactions. 
These were completed \cite{jlms07,djlss08, kjlms09}
 with the results on the total $\pi N$ cross sections shown in
Fig. \ref{fig:pin}. The model was then applied
to analyze pion photoproduction \cite{jlmss08} and 
electroproduction \cite{jklmss09} data. 
An important feature of the DCC model is the possibility
to distinguish 
 the $N^*$ excitation due to the quark-gluon core from that due to
the meson-cloud (meson-baryon dressing) effects. This is illustrated 
in Fig. \ref{fig:delta} for the $G^*_M$  form factor of the 
$\gamma N \rightarrow \Delta$(1232)P$_{33}$ transition. It is further
found \cite{Sato,jlmss08} that the contribution due to the quark-gluon core is
close to the prediction of constituent quark models. 

\begin{figure}[t]
\vspace{10pt}
\begin{center}
\mbox{\epsfig{file=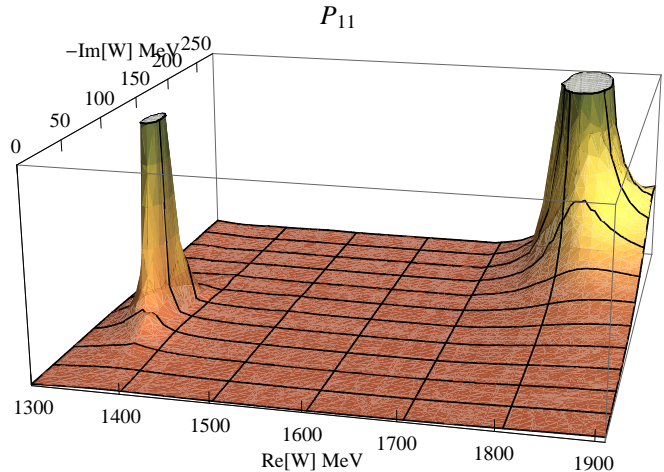, width=9cm}}
\end{center}
\caption{The $P_{11}$ resonance poles 
extracted \cite{ssl09} from the
dynamical coupled-channel model of Refs. \cite{jlms07,kjlms09}.
}
\label{fig:pole}
\end{figure}

In parallel, an exact analytic continuation method for extracting nucleon 
resonances from multi-channel reactions has been 
developed \cite{ssl09} and applied \cite{sjklms09} to
extract the nucleon resonances from this model. 
As an example, Fig. \ref{fig:pole} shows the positions of the extracted $P_{11}$
resonances in the
complex energy plane. Within the DCC model, it is found that these
two $P_{11}$ resonances evolve from the same bare state
with a mass of about 1700 MeV, which could be identified with the
hadron structure calculations in the absence of meson-baryon degrees of freedom,
such as the constituent quark model.

The plan at EBAC 
is to complete the analysis of single $\pi$ and $\eta$ production up
to $W= 2$ GeV and $Q^2 \leq$ 6 GeV$^2$,
to analyze the 2$\pi$ production data, and
to extract the $N^*$ form factors at the
resonance poles.
With these, EBAC will complete its first-stage analysis in 2010.

EBAC's second-stage work is to analyze the data of $K\Lambda$
and $K\Sigma$ production.
EBAC is collaborating with the CLAS collaboration to
develop methods for extracting $\gamma N \rightarrow K\Lambda$ multipole
amplitudes as model independently as possible from the forthcoming JLab
data from the over-complete measurements of $K\Lambda$ 
and $K\Sigma$ production.

Efforts are also being made to develop interpretations of
the extracted $N^*$ parameters in terms of hadron structure calculations.
In addition to considering constituent quark models, we
are also investigating
how LQCD results can be related to the dynamical coupled-channel analysis
\cite{yl09}.

\section{Future prospects of the $N^*$ program at 6  GeV and with the 12-GeV
upgrade.}

The main part of the experimental program to search for predicted new baryon
states
lies still in the future. The polarized frozen spin target facility FROST
will be installed
in CLAS in 2010 to complete the extensive program to measure double and
triple
polarization observables with protons that are polarized transversely to the
production
plane and to use linearly and circularly polarized photons. 
For the kaon-hyperon
final states,
the recoil polarization will be measured as well.  Data taking with
FROST will
be followed with the equivalent program on polarized neutrons. This part
will use the
polarized HD target and provide complementary information on the isospin
content of
the reaction and of the excited state. This information is required to
separate isospin 1/2
from isospin 3/2 excitations. Moreover, selection rules allow the
photo-excitation of
some states, e.g. N(1675)D$_{15}$, from neutrons, but not from 
protons.

With the 12-GeV upgrade, the $N^*$ program at JLab is being extended as an 
essential part of the comprehensive program of exclusive electroproduction
measurements with the CLAS12 detector. A proposal for measuring $\pi^0$, $\eta$,
and charged multi-pion final states in the unexplored domain of $Q^2$,
from 5.0 to 12 GeV$^2$, has recently been approved by JLab's Program Advisory
Committee of 2009.

The proposed experiments have three objectives \cite{theory_support}. 
The first is to map out the quark structure of $N^*$s from the 
data on exclusive meson electroproduction reactions. 
In the considered  $Q^2 > 5.0$~GeV$^2$ domain,
the meson-baryon dressing is weak, as illustrated
in Fig. \ref{fig:delta} for the $\Delta$(1232)P$_{33}$, and hence the extracted
$\gamma^* N \rightarrow N^*$ form factors
can be used more directly to probe the quark
substructure of $N^*$s.

The second objective is to investigate the dynamics of dressed quark
interactions inside the nucleon core and to understand how these
interactions emerge from QCD. We are motivated by the recent advance in
developing hadron models based on the
Dyson-Schwinger equations (DSE) of QCD~\cite{Holl,Eich},
as well by exploratory attempts to evaluate transition
$\gamma^*NN^*$ from factors starting from QCD Lagrangian within the
framework of LQCD \cite{Alexandrou,Braun,AlexBeij,Lin}. 

The third objective is to study the $Q^2$-dependence of the non-perturbative 
dynamics of QCD. This is based on the recent investigation of the momentum 
dependence of the dressed quark mass of the quark propagator 
within LQCD~\cite{mass-lqc} and DSE~\cite{Bhag}. 
Our focus is on the important question of  
how baryon structure emerges from confinement and dynamical chiral 
symmetry breaking of QCD.

The proposed experiments are closely related to 
the GPD program at JLab. The specification of exclusive reactions
at high momentum transfer in terms of GPDs is a major goal of the
CLAS12 upgrade. Experiments that are already approved as part of this program
include  deeply virtual Compton scattering and deeply  virtual
meson production.


\end{document}